\documentclass[12pt]{article}
\usepackage{amsmath}
\usepackage{graphicx,psfrag,epsf}
\usepackage{enumerate}
\usepackage{natbib}
\usepackage{url} % not crucial - just used below for the URL
\usepackage{placeins}
\usepackage{color}
\newcommand{\blind}{0}

\newcommand{\pkg}[1]{{\normalfont\fontseries{b}\selectfont #1}} 
\let\proglang=\textit
\let\code=\texttt

\usepackage{mathtools,amssymb,amsthm}
\usepackage{bm}
\usepackage[linesnumbered,ruled,vlined]{algorithm2e}
\newtheorem{result}{Result}
\newtheorem{theorem}{Theorem}

\newcommand{\bB}{\bm{B}}

\newcommand{\bD}{\bm{D}}
\newcommand{\bX}{\bm{X}}
\newcommand{\bI}{\bm{I}}
\newcommand{\bU}{\bm{U}}
\newcommand{\bV}{\bm{V}}

\newcommand{\bx}{\bm{x}}
\newcommand{\bb}{\bm{b}}

\newcommand{\bK}{\bm{K}}

\newcommand{\by}{\bm{y}}
\newcommand{\bY}{\bm{Y}}
\newcommand{\bc}{\bm{c}}
\newcommand{\bk}{\bm{k}}

\newcommand{\bz}{\bm{z}}
\newcommand{\be}{\bm{e}}
\newcommand{\br}{\bm{r}}
\newcommand{\bv}{\bm{v}}
\newcommand{\bu}{\bm{u}}
\newcommand{\bmu}{\bm{\mu}}

\newcommand{\btht}{\bm{\theta}}

\newcommand{\hbt}{\hat{\bm{\theta}}}
\newcommand{\hbT}{\hat{\bm{\Theta}}}
\newcommand{\bxi}{\bm{\xi}}
\newcommand{\bLam}{\bm{\Lambda}}

\newcommand{\bSig}{\bm{\Sigma}}

\newcommand{\tr}{\text{tr}}

\newcommand{\tE}{\text{E}}
\newcommand{\hbc}{\hat{\bm{c}}}
\newcommand{\Var}{\text{Var}}
\newcommand{\tvec}{\text{vec}}
\newcommand{\nsimexp}{three }

\newcommand{\hsigsq}{\hat{\sigma}^2}
\newcommand{\hsig}{\hat{\sigma}}
\newcommand{\tsigsq}{\tilde{\sigma}^2}
\newcommand{\tsig}{\tilde{\sigma}}

\newcommand{\ttr}{\text{tr}}

\usepackage{chngcntr}

\begin{document}

\def\spacingset#1{\renewcommand{\baselinestretch}%
{#1}\small\normalsize} \spacingset{1}

%%%%%%%%%%%%%%%%%%%%%%%%%%%%%%%%%%%%%%%%%%%%%%%%%%%%%%%%%%%%%%%%%%%%%%%%%%%%%%

\if0\blind { \title{\bf A Sequential Design Approach for Calibrating a Dynamic Population Growth Model}
  \author{Ru Zhang, C. Devon Lin\hspace{.2cm}\\
    Department of Mathematics and Statistics, \\
    Queen's University, ON, Canada\\
    and\\
    Pritam Ranjan\\
    Operations Management \& Quantitative Techniques, \\
    Indian Institute of Management Indore, MP, India} \date{}
  \maketitle
} \fi

\if1\blind
{
  \bigskip
  \bigskip
  \bigskip
  \begin{center}
    {\LARGE\bf Title}
\end{center}
  \medskip
} \fi

\bigskip
\begin{abstract}

A comprehensive understanding of the population growth of a variety of pests is often crucial for efficient crop management. Our motivating application comes from calibrating a two-delay blowfly (TDB) model which is used to simulate the population growth of \emph{Panonychus ulmi} (Koch) or European red mites that infest on apple leaves and diminish the yield. We focus on the inverse problem, that is, to estimate the set of parameters/inputs of the TDB model that produces the computer model output matching the field observation as closely as possible. The time series nature of both the field observation and the TDB outputs makes the inverse problem significantly more challenging than in the scalar valued simulator case.

In spirit, we follow the popular sequential design framework of computer experiments. However, due to the time-series response, a singular value decomposition based Gaussian process model is used for the surrogate model, and subsequently, a new expected improvement criterion is developed for choosing the follow-up points. We also propose a new criterion for extracting the optimal inverse solution from the final surrogate. Three simulated examples and the real-life TDB calibration problem have been used to demonstrate higher accuracy of the proposed approach as compared to popular existing techniques. 

\end{abstract}

\noindent%
{\it Keywords:} Computer model, Expected improvement, Gaussian process model, Option pricing,
Saddlepoint approximation, Singular value decomposition, Time series.
\vfill

\newpage
\spacingset{1.45} % DON'T change the spacing!

\section{Introduction}
\label{sec:intro}

The motivating application comes from the apple farming industry in the Annapolis Valley, Nova Scotia, Canada, in which the underlying objective is to have a comprehensive understanding of the population growth of a particular variety of pest called the European red mites (ERM) or \emph{Panonychus ulmi} (Koch). ERM were originally brought to North America from Europe, and they have been playing an important role in controlling the abundance of different species of pests. However, it is common that  ERM infest on apple leaves, resulting in poor yields, and hence a  concern for apple farmers in the Annapolis Valley. Among others, \cite{teismann2009} and \cite{franklin_2014} use mathematical-biological models based on predator-prey dynamics to gain cheaper insights in the ERM population growth over a season. Such models are referred to as the {\em computer models} or {\em simulators} in the computer experiments literature.

The computer model by \cite{franklin_2014} is a differential equation based Two-Delay Blowfly (TDB) model, which consists of eleven parameters treated as the inputs to the model, and produces time-series outputs that characterize the ERM population growth. Let $\by(\bx)=[y_{t_1}(\bx),\dots,y_{t_L}(\bx)]^T\in\mathbb{R}^L$, denote the model output at input $\bx\in\Omega\subseteq\mathbb{R}^q$, and $L$ time points $t_1< \dots < t_L$. The present research endeavour is motivated by the task of solving an inverse problem, i.e., finding the best possible inputs of the population growth model to produce realistic outputs (i.e., closer to the field data collected on ERM population count from apple trees, denoted by $\bxi=[\xi_{t_1},\dots,\xi_{t_L}]^T\in\mathbb{R}^L$).

In computer experiments, computer models are popular substitutes for gaining insights into complex processes that are often too expensive, time-consuming or sometimes even infeasible to observe. The applications range from medical, agricultural, industrial, cosmological to climate sector, etc., and the scientific goals include process optimization, estimating a pre-specified process feature, an overall understanding of the phenomena and identification of important variables.  The computer models for complex processes are also sometimes computationally demanding and/or with very high input dimensions. In such a scenario, it is crucial to build
an emulator that acts as a proxy for the simulator,  for searching the optimal solution. Gaussian process (GP) based emulators are popular, inexpensive surrogates for computer models (i.e., simulators), and a sequential design approach often leads to efficient selection of inputs for accurate estimation of the feature of interest.

The inverse problem introduced here is also sometimes referred to as the calibration of the computer model. It shall be noted, however, that the term ``model calibration" has other connotations as well in computer experiments literature. We further assume that the field observation contains some random noise, but the simulator (i.e., the computer model) is deterministic without any systematic bias/discrepancy with respect to the field observation \citep{kennedy2001bayesian, higdon2008computer, pratola2013fast}. Let $\delta(\bx) = \|\bxi-\by(\bx)\|_2^2$ be the discrepancy between the target response (i.e., the field observation) and the model output, then the inverse problem refers to locating an input $\bx^*$ in the design domain $\Omega$ such that
\begin{align}\label{eq:ls-sol}
  \bx^*=\underset{\bx\in\Omega}{\mathrm{argmin}}\ \delta(\bx),
\end{align}
where $\|\cdot\|_2$ is the $L_2$ norm.

The inverse problem for expensive to evaluate scalar-valued computer models has been extensively studied over the last decade (e.g., \cite{ranjan2008sequential, picheny2013quantile}; Bingham et al. (2014)). However, the inverse problem for dynamic computer simulators has not been explored in depth. A few attempts include, a history matching algorithm by \cite{vernon2010galaxy} and \cite{andrianakis2017efficient} - simultaneously solves a handful of scalar valued inverse problems, and the direct emulation and optimization of the scalarized discrepancy $\delta(\bx)$ by \cite{pratola2013fast} and \cite{ranjan2016inverse}.

In this paper, we present a novel approach for  the inverse problem of dynamic computer simulators. We adopt the singular value decomposition (SVD)-based GP models and the empirical Bayesian inference procedure proposed  by \citep{zhang2017local} for emulating the simulator response. We take up an expected improvement (EI) criterion-based sequential strategy promoted by \cite{jones1998efficient} that was originally used to response optimization. Our  main contributions are two-folded: (a) a new EI criterion is proposed for searching for the next design point; and (b) a new criterion for extracting the solution of the inverse problem from the final surrogate.  The new EI criterion does not admit a closed form, and we use the saddlepoint approximation proposed by  \cite{huang2011saddlepoint} for its evaluation.  To apply the approximation, necessary statistical quantities are derived and interesting theoretical results are proven.  We have also implemented all important functions and algorithms in the \proglang{R} package \pkg{DynamicGP}.

The remaining sections are arranged as follows. Section \ref{sec:mot-app} describes  the motivating application in details.  Section \ref{sec:svd-gpm} formulates the inverse problem for dynamic computer
experiments and reviews the SVD-based GP model as a statistical
emulator for dynamic computer simulators.  In Section
\ref{sec:new-method}, we introduce the saddlepoint approximation based EI criterion for the inverse problem and a new approach for
extracting the solution. Section \ref{sec:sim-study} uses three simulated examples and a real application to illustrate the advantages of the proposed method compared with the two existing
alternatives. Concluding remarks are provided in Section
\ref{sec:con-rmk}. All the proofs are relegated to Appendices.

\section{Motivating Application}
\label{sec:mot-app}

One variety of pest common among tree fruit crops, including apple trees, is \textit{Panonychus ulmi} (Koch) or European red mites (ERM), which first appeared in North America in the early 1900's. ERM survive by consuming photosynthetic cells found in the leaves. Interestingly, several studies have shown that ERM are not primary pests as their abundance typically does not reach sustainable levels due to predator-prey dynamics \citep{franklin_2014}. However, if their predators are extinguished through the use of pesticides, ERM become the primary pest which may lead to infestation and hence decreased photosynthesis. This in turn causes browning of the apple tree leaves, softer apples, premature fruit drops, and lower yields for the farmer. Consequently, a comprehensive understanding of the ERM population dynamics is crucial for effective crop management.

Annapolis Valley in Nova Scotia, Canada, is famous for a wide variety and superior quality apples. Unfortunately, infestation due to ERM is a serious concern for these farmers, who are often forced to rely on expensive pesticides in order to limit the severity that pest infestation may have on their crops. However, these pesticides can have detrimental effects on neighbouring organisms, and over the years, the Canadian government has funded numerous research projects aimed at finding alternative methods of pest control \citep{getz_1982, hardman_1991}.

ERM start their lives as \textit{winter eggs} that are laid in the late summer months of the previous year. Once the temperature rises to a sufficient level in the following spring, winter eggs hatch and emerge as larvae. From there, ERM grow into two broad groups: non egg-laying \textit{juveniles} and egg-laying \textit{adults}. During the summer, the adult female ERM lay \textit{summer eggs} that are significantly softer than their winter counterparts and hatch during the same season due to the warmer climate. Finally, in mid-to-late August, ERM transition from laying summer eggs to winter eggs and the cycle repeats itself \citep{parent_1957}.

Actual field data on the population of different stages of ERM growth cycle is difficult to collect. For sampling purposes, the orchards have to be divided in plots, plots into trees, and all the way down to leaves. Finally, different forms of ERM (eggs, larvae, juvenile, and adults) are counted off randomly selected sets of leaves. For more economical insights, several biological/mathematical models have been proposed thus far \citep{hardman_1998, franklin_2014}. In this paper, we focus on a Two-Delay Blowfly (TDB) model which is an adaptation of the popular Blowfly delay differential equation, and introduces separate delay parameters for each of the two major life stages, summer egg to juvenile, and juvenile to adult. The TDB model takes an eleven-dimensional input (calibration parameters), and produces three time-series outputs representing the population growth of summer eggs, juveniles and adults. These calibration parameters include death rates, fecundity, delay parameters, initial population, and so on. Some of these parameters are based on ecological principles which make it suitable for population growth model under different climates.

\begin{figure}[h!]
  \centering
  \includegraphics[width=6.0in]{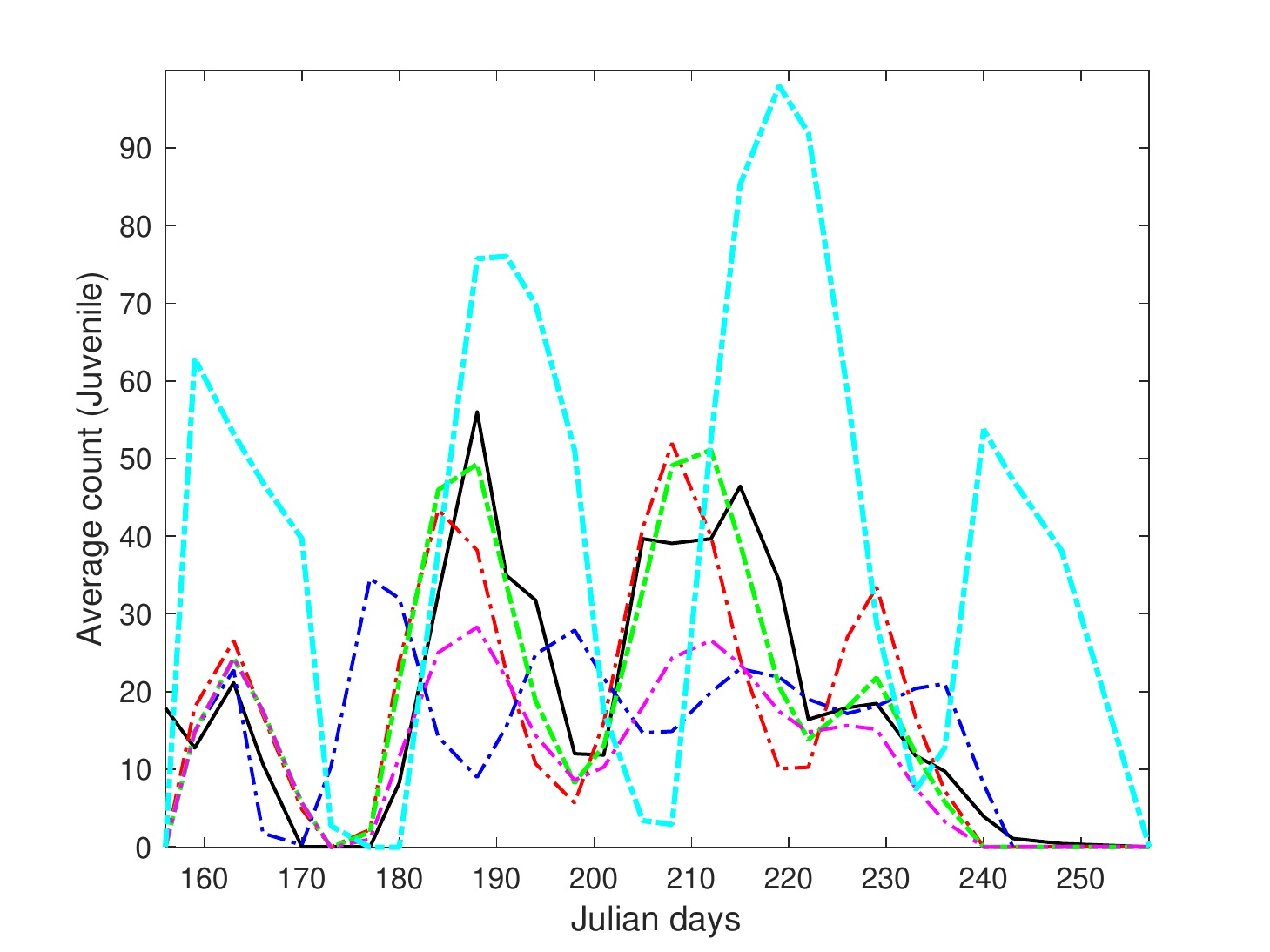}
  \caption{Juvenile ERM population dynamics as outputs of the TDB model at five different inputs. The solid curve shows the target response, and the dashed curves show the TDB outputs.}
  \label{fig:tdb-show}
\end{figure}

As in \cite{franklin_2014}, our scientific objective is to find the optimal set of inputs for the TDB model that gives a good approximation to a pre-specified target. The intent behind this inverse problem is to calibrate the TDB model to produce realistic outputs closer to the reality. In this paper, we discuss the calibration of the TDB model with respect to only the juvenile population. The target response corresponds to the average count of juvenile population over a 28-day sample during 156-th to 257-th Julian days. Figure~\ref{fig:tdb-show} presents the target response (field data) and a few randomly generated TDB runs.

%\begin{comment}
%\end{comment}
%

The existing methodologies to solve this inverse problem start by finding the discrepancy $\delta(\bx) = \|\bxi - \by(\bx)\|$ between the target response $\bxi$ and the TDB runs $\by(\bx)$ for every input $\bx$ in the input space, and then find the minimizer of $\delta(\bx)$ by either some brute-force approach \citep{franklin_2014} or an efficient algorithm based on a sophisticated statistical modeling \citep{pratola2013fast,ranjan2016inverse}. In this paper, we develop a new methodology that uses the dynamic (time-series) structure of the response to solve the underlying inverse problem more efficiently, i.e., require fewer TDB runs to achieve the desired accuracy.

\section{SVD-based GP Model}
\label{sec:svd-gpm}

The dynamic computer simulator $\by(\cdot)$ may be expensive to evaluate and/or take high-dimensional inputs, which precludes using the numerical optimization algorithms on simulator runs to search for optimal $\bx^*$ in (\ref{eq:ls-sol}). As a computationally efficient alternative, an emulator can approximate the simulator response and accurately quantify the uncertainty of the simulator response at an arbitrary untried point in the design domain, which can be used further for a focussed search of the inverse solution. This section briefly reviews the SVD-based GP model \citep{higdon2008computer} as an emulator for dynamic computer simulators and its empirical Bayesian inference \citep{zhang2017local}.

Suppose the responses of the computer simulator $\by(\cdot)$ have been collected at 
$N$ design points, where $\bm{X}=[\bm{x}_1,\dots,\bm{x}_N]^T$ be the $N\times q$ design matrix
and $\bm{Y}=[\bm{y}(\bm{x}_1),\dots,\bm{y}(\bm{x}_N)]$ be the
$L\times N$ matrix of time series responses. The SVD on $\bm{Y}$ gives
$\bm{Y}=\bm{U}\bm{D}\bm{V}^T$, where
$\bm{U}=[\bm{u}_1,\dots,\bm{u}_{k}]$ is an $L\times k$
column-orthogonal matrix, $\bm{D}=\text{diag}(d_1,\dots,d_{k})$ is a
$k \times k$ diagonal matrix of singular values sorted in decreasing
order, $\bm{V}$ is an $N\times k$ column-orthogonal matrix of right
singular vectors, and $k=\min\{N,L\}$. The SVD-based GP model assumes
that, for any $\bx\in\mathbb{R}^q$,
\begin{align}\label{eq:md}
\bm{y}(\bm{x})=\sum_{i=1}^pc_i(\bm{x})\bm{b}_i+\bm{\epsilon},
\end{align}
where the orthogonal basis $\bm{b}_i=d_i\bm{u}_i\in \mathbb{R}^L$, for
$i=1,\dots,p$. The coefficients $c_i$'s in (\ref{eq:md}) are assumed
to be independent Gaussian processes, i.e.,
$c_i\sim \mathcal{GP}(0,\sigma^2_iK_i(\cdot,\cdot;\btht_i))$ for
$i=1,\dots,p$, where $K_i$'s are correlation functions. We use the
popular anisotropic Gaussian correlation,
$K(\bm{x}_1,\bm{x}_2;\btht_i)=\exp\{-\sum_{j=1}^q\theta_{ij}(x_{1j}-x_{2j})^2\}$
\citep{santner2003design}. The residual term $\bm{\epsilon}$ in
(\ref{eq:md}) is assumed to be independent $\mathcal{N}(0,\sigma^2\bI_L)$.

The number of significant singular values, $p$, in (\ref{eq:md}), is
determined empirically by the cumulative percentage criterion
$p=\min\{m:(\sum_{i=1}^md_i)/(\sum_{i=1}^{k}d_i)>\gamma\}$,
where $\gamma$ is a threshold of the explained variation. We use
$\gamma=95\%$ in this article.

For all the model parameters in (\ref{eq:md}), we use the maximum a
posteriori (MAP) values as the plug-in estimates. To obtain the MAP
estimates, we impose inverse Gamma priors for the process and noise
variances $\sigma^2_i$ and $\sigma^2$ \citep{gramacy2015local}, i.e.,
\begin{align*}%\label{eq:prior}
  \begin{aligned}
    [\sigma_i^2]\sim\text{IG}\left(\frac{\alpha_i}{2},\frac{\beta_i}{2}\right),
    i=1,\dots,p,
  \end{aligned} \qquad
  \begin{aligned} [\sigma^2]\sim \text{IG}\left(\frac{\alpha}{2},\frac{\beta}{2}\right),
  \end{aligned}
\end{align*}
and use the Gamma prior for the hyper-parameter $1/\theta_{ij}$
\citep{gramacy2015lagp}.

It is shown by \cite{zhang2017local} that the approximate predictive
distribution produced by model (\ref{eq:md}) for an arbitrary untried
$\bx_0\in\mathbb{R}^q$ is obtained by
\begin{align}\label{eq:final}
\pi(\bm{y}(\bm{x}_0)|\bm{Y})
\approx \pi(\bm{y}(\bm{x}_0)|\bm{V}^*,\hat{\bm{\Theta}},\hat{\sigma}^2) \approx
\mathcal{N}\big(\bm{B}\hat{\bm{c}}(\bm{x}_0|\bV^*,\hat{\bm{\Theta}}),\bB\bm{\Lambda}(\bV^*,\hat{\bm{\Theta}})\bB^T+\hat{\sigma}^2\bI_L\big),
\end{align}
where $\bB= [d_1\bu_1,\dots,d_p\bu_p] = \bm{U}^*\bm{D}^*$, with $\bm{U}^*=[\bm{u}_1,\dots,\bm{u}_p]$, $\bm{D}^*=\text{diag}(d_1,\dots,d_p)$ and $\bV^*=[\bv_1,\dots,\bv_p]^T$, and
$\hbT=\{\hbt_1,\dots,\hbt_p\}$ and $\hat{\sigma}^2$ are the MAP estimates of the
correlation parameters and noise
variance $\sigma^2$, respectively. As shown in \cite{zhang2017local},
\begin{align}\label{eq:resid-var}
  \hbt_i=\underset{\btht_i\in\mathbb{R}^q}{\mathrm{argmax}}\:|\bK_i|^{-1/2}\left(\frac{\beta_i+\psi_i}{2}\right)^{-(\alpha_i+N)/2}\pi(\btht_i),
  \quad \text{and} \quad \hat{\sigma}^2=\frac{1}{NL+\alpha+2}\left(\br^T\br+\beta\right),
\end{align}
where $\bm{K}_i$ is the $N\times N$ correlation matrix on the design
matrix $\bm{X}$ with the $(j,l)$th entry being
$K(\bm{x}_j,\bm{x}_l;\hbt_i)$ for $i=1,\dots,p$ and $j,l = 1,\dots,N$,
$\psi_i=\bv_i^T\bK_i^{-1}\bv_i$, $\pi(\btht_i)$ is the prior
distribution of $\btht_i$ and
$\br=\tvec(\bY)-(I_N\otimes\bB)\tvec(\bV^{*T})$ with $\tvec(\cdot)$
and $\otimes$ being the vectorization operator and the Kronecker
product for matrices, respectively.

The vector of predictive mean of the coefficients at $\bx_0$
is
\begin{align}\label{eq:coef-pmean}
\hbc(\bx_0,|\bV^*,\hbT)=[\hat{c}_1(\bx_0|\bv_1,\hbt_1),\dots,\hat{c}_p(\bx_0|\bv_p,\hbt_p)]^T=[\bk_1^T(\bx_0)\bK_1^{-1}\bv_1,\dots,\bk_p^T(\bx_0)\bK_p^{-1}\bv_p]^T,
\end{align}
where
$\bm{k}_i(\bm{x}_0)=[K(\bm{x}_0,\bm{x}_1;\hbt_i),\dots,K(\bm{x}_0,\bm{x}_N;\hbt_i)]^T$.
The predictive variance $\bm{\Lambda}(\bV^*,\hat{\bm{\Theta}})$ of the
coefficients at $\bx_0$ is a $p\times p$ diagonal matrix with the
$i$th diagonal entry being
\begin{align}\label{eq:coef-ps2}
  \hat{\sigma}_i^2(\bm{x}_0|\bv_i,\hat{\bm{\theta}}_i)=\frac{(\beta_i+\bv^T_i\bK_i^{-1}\bv_i)\left(1-\bm{k}^T_i(\bm{x}_0)\bm{K}^{-1}_i\bm{k}_i(\bm{x}_0)\right)}{\alpha_i+N}.
\end{align}

The built-in function called \code{svdGP} in the \proglang{R} package \pkg{DynamicGP} provides an easy implementation of this surrogate model \citep{dynamicGP}. The arguments of \code{svdGP} can also be tuned to speed up the computation by parallelization.

\section{New Methodology}
\label{sec:new-method}

Restricted by the computational cost of the dynamic computer simulator, the budget for the number of the computer simulator evaluations may be very limited. In such a scenario, sequential design frameworks are efficient in estimating a pre-specified feature of interest (FOI) with high accuracy.

The procedure starts by choosing a space-filling design $\bX$ of size $n_0$ from the input space, and then evaluate the computer simulator (TDB model, in our application) at the training points to produce the response matrix $\bY$. Subsequently, the training sets $\bX$ and $\bY$ are used to fit the SVD-based GP model, a statistical surrogate of the computer simulator. One of the most important aspects of the sequential design framework is to carefully select the follow-up point that leads to further improvement in the current best estimate of FOI. In computer experiments, the expected improvement (EI) criterion is considered as the gold standard for finding efficient designs (see for example, \cite{jones1998efficient} and Bingham et al. (2014)). Once the follow-up point is chosen, the simulator output is obtained for this new point, and the surrogate fit is updated. These two steps, choosing follow-up point by optimizing EI criterion and updating the surrogate, are repeated until either the budget of the follow-up points (say, $n_{new}$) runs out or the FOI estimate has stabilized. At the end, the best estimate of the FOI is extracted from the final surrogate fit. In this section, we focus on deriving an appropriate EI criterion under the dynamic computer model setup.

\subsection{EI Criterion for Follow-up Design Points}
\label{sec:seq-des}

To efficiently choose follow-up design points of addressing the calibration/inverse problem (\ref{eq:ls-sol}), we propose an EI criterion. Our objective  is to locate the optimal set of inputs $\bx\in\Omega$ such that $\delta(\bx) = \|\bxi-\by(\bx)\|^2_2$ is minimized. Within the current design set $\bX=\{\bx_1,...,\bx_N\}$, the minimum squared $L_2$ discrepancy is
\begin{align*}
  \delta_{\min}=\min_{i=1,\dots,N} \delta(\bx_i),
\end{align*}
where $\bx_i$ is the $i$th design point in $\bX$, for
$i=1,\dots,N$. We intend to search for the next design point that
improves the currently attained $\delta_{\min}$ most. To achieve this
objective, we follow the sequential design framework in \cite{jones1998efficient} and define the improvement function
\begin{align}\label{eq:improve-fun}
I(\bx)=\big(\delta_{\min}-\delta(\bx)\big)_{+},
\end{align}
where $(u)_{+}=\max\{0,u\}$ for $u\in\mathbb{R}$. The improvement
function represents the decrease of the squared $L_2$ discrepancy at
$\bx$ with respect to $\delta_{\min}$. In the spirit of the EI formulation in \cite{jones1998efficient}, we define the EI criterion as
\begin{align}\label{eq:ei-crit}
  E[I(\bx)]=\tE\Big[\big(\delta_{\min}-\delta(\bx)\big)_{+}\Big|\bY\Big],
\end{align}
\noindent where the expectation is taken with respect to the predictive distribution of $\delta(\bx)$ given $\bY$. Under the SVD-based GP surrogate model in (\ref{eq:md})-(\ref{eq:final}), the EI criterion (\ref{eq:ei-crit}), in general, does not have a closed form expression if the expectation is taken with respect to the predictive distribution of $\delta(\bx)$ or $\by(\bx)$ given $\bY$. Consequently, we focus on finding a good approximation of (\ref{eq:ei-crit}) in this section.

A few conventional approaches to approximate such an expectation include Monte Carlo integration, Simpson’s rule and Gaussian quadrature, however, these numerical methods are computationally too intensive to achieve satisfactory accuracy level \citep{robert2005monte, davis2007methods,roshan_uncertainty}. Recently, \cite{huang2011saddlepoint} developed saddlepoint approximations of $E[(Z-K)_+]$, where $Z$ is a sum of independent random variables and $K$ is a constant. This expectation can be viewed as the payoff in the context of option pricing, stop-loss premium or expected shortfall of an insurance portfolio. We apply this method to find a closed form expression for the so-called saddlepoint approximation-based expected improvement (saEI) criterion for (\ref{eq:ei-crit}). In the Supplementary Material, we present an empirical study which demonstrates that the saddlepoint approximation method gives high accuracy in much less computational time as compared to the Monte Carlo integration for evaluating the EI criterion in (\ref{eq:ei-crit}).

The saddlepoint approximation method in \cite{huang2011saddlepoint} uses some distributional features of $\delta(\bx)$, in particular, the expected discrepancy and some statistics derived from its cumulant generating function, $\kappa_{\delta}(s)$ (in short for $\kappa_{\delta(\bx)}(s)$). The cumulant generating function is the log- moment generating function ($M_{\delta}$), i.e., $\kappa_{\delta}(s) = \log(M_{\delta}(s))$. For simplicity in notation, we drop the dependency on $\bx$ and conditioning on variables and parameters in the following discussion. Let $\bz=\bxi-\by(\bx)$, then the predictive distribution of $[\bz|\bY]$ is $\mathcal{N}(\bmu,\hat{\bSig})$, with $\bmu = \bxi-\bB\hat{\bc}$, $\hat{\bSig}=\bB\bm{\Lambda}\bB^T+\hat{\sigma}^2\bI_L$, and the moment generating function of $\delta(\bx) = \bz^T\bz$ is 
\begin{align*}
  M_{\delta}(s)
  =(2\pi)^{-\frac{L}{2}}|\hat{\bSig}|^{-\frac{1}{2}}\exp\left\{-\frac{\bmu^T\hat{\bSig}^{-1}\bmu}{2}\right\}\int_{\mathbb{R}^L}\exp\left\{-\frac{\bz^T(\hat{\bSig}^{-1}-2s\bI_L)\bz-2\bmu^T\hat{\bSig}^{-1}\bz}{2}\right\}d\bz.
\end{align*}
Following Lemma B.1.1 of \cite{santner2003design}, if $\hat{\bSig}^{-1}-2s\bI_L$ is positive definite, the above expression of $M_{\delta}(s)$ can be simplified to 
\begin{align*} 
  M_{\delta}(s)=|\bI_L-2s\hat{\bSig}|^{-\frac{1}{2}}\exp\left\{\frac{\bmu^T\hat{\bSig}^{-1}(\hat{\bSig}^{-1}-2s\bI_L)^{-1}\hat{\bSig}^{-1}\bmu-\bmu^T\hat{\bSig}^{-1}\bmu}{2}\right\},
\end{align*}
and subsequently, the desired cumulant generating function of $\delta(\bx)$ is
\begin{align}\label{eq:cum-out}
  \kappa_{\delta}(s)={-\frac{1}{2}}\log(|\bI_L-2s\hat{\bSig}|) + \frac{1}{2}\ \bmu^T\hat{\bSig}^{-1}(\hat{\bSig}^{-1}-2s\bI_L)^{-1}\hat{\bSig}^{-1}\bmu - \frac{1}{2}\ \bmu^T\hat{\bSig}^{-1}\bmu .
\end{align}

The closed form expressions of the expected discrepancy and the first three derivatives of the cumulant generating function are presented here.
\begin{result}\label{result:mu_delta}
The expected discrepancy with respect to the predictive distribution (\ref{eq:final})  is given by 
\begin{align}
  \mu_\delta &= \tE[\delta(\bx)|\bY]  =\bxi^T (\bI_L - \bU^*\bU^{*T})\bxi  + \sum_{i=1}^pd_i^2\left[(\hat{c}_i(\bx)-\hat{c}_{\bxi,i})^2 + \hat{\sigma}^2_i(\bx)\right] +\hat{\sigma}^2L. \label{eq:mu-delta}
\end{align}
where $\hat{\bc}_{\bxi}=\bD^{*-2}\bB^T\bxi$.
\end{result}

\begin{result}\label{result:cum-gen-fn_final}
The cumulant generating function in (\ref{eq:cum-out}) can be simplified to 
\begin{align}
  \kappa_{\delta}(s)=&-\frac{L-p}{2}\log(1-2s\hat{\sigma}^2)-\frac{1}{2}\sum_{i=1}^p\log(1-2s(\hat{\sigma}^2+\hat{\sigma}^2_id_i^2))\\ \nonumber
  &+\sum_{i=1}^p\frac{2s^2\hat{\sigma}_i^2 \mu^2_{bi}}{[1-2s(\hat{\sigma}_i^2d_i^2+\hat{\sigma}^2)](1-2s\hat{\sigma}^2)} + \frac{s\bmu^T\bmu}{1-2s\hat{\sigma}^2},
\end{align}
where $\mu_{bi}$ is the $i$-th entry of $\bB^T\bmu$.
\end{result}

\begin{result}\label{result:cum-gen-derivative}
Let $\tilde{\sigma}_i^2=\hat{\sigma}_i^2d_i^2+\hat{\sigma}^2$. Then, $\kappa^{(i)}_{\delta}(s)$, the $i$-th order derivative of $\kappa_{\delta}(s)$ with respect to $s$, for $i=1,2,3$, are as follows:
\begin{align*}
  \kappa^{(1)}_{\delta}(s)=&\frac{\bmu^T\bmu}{(1-2s\hsigsq)^2}+(L-p)\frac{\hsigsq}{1-2s\hsigsq}+\sum_{i=1}^p\frac{\tsigsq_i}{1-2s\tsigsq_i}\\
&+\sum_{i=1}^p\frac{4\hsigsq_i\mu_{bi}^2\ s(1-s\tsigsq_i-s\hsigsq)}{(1-2s\tsigsq_i)^2(1-2s\hsigsq)^2},
\end{align*}
\begin{align*}
  \kappa^{(2)}_{\delta}(s)=&\frac{4\hsigsq\bmu^T\bmu}{(1-2s\hsigsq)^3}+(L-p)\frac{2\hsig^4}{(1-2s\hsigsq)^2}+\sum_{i=1}^p\frac{2\tsig_i^4}{(1-2s\tsigsq_i)^2}\\
&+\sum_{i=1}^p\frac{4\hsigsq_i\mu_{bi}^2\ (1-12s^2\tsigsq_i\hsigsq+8s^3\tsig_i^4\hsigsq+8s^3\tsigsq_i\hsig^4)}{(1-2s\tsigsq_i)^3(1-2s\hsigsq)^3},
\end{align*}
\begin{align*}
  \kappa^{(3)}_{\delta}(s)=&\frac{24\hsig^4\bmu^T\bmu}{(1-2s\hsigsq)^4}+(L-p)\frac{8\hsig^6}{(1-2s\hsigsq)^3}+\sum_{i=1}^p\frac{8\tsig_i^6}{(1-2s\tsigsq)^3}\\
&+\frac{24\hsigsq_i\mu_{bi}^2\ (\tsigsq_i+\hsigsq-8s\tsigsq_i\hsigsq+32s^3\tsig_i^4\hsig^4-16s^4\tsig^6_i\hsig^4-16s^4\tsig_i^4\hsig^6)}{(1-2s\tsigsq_i)^4(1-2s\hsigsq)^4}.\\
\end{align*} 
\end{result}

Appendix~A presents the proofs of Results~1 and 2. Result~3 is a straightforward application of differentiation of $\kappa_{\delta}(s)$ with respect to $s$ and does not require an explicit proof. Now, the saddlepoint approximation method begins by finding a solution of the derivative equation,
\begin{align}\label{eq:saddle-eq}
  \kappa_{\delta}^{(1)}(s)=\delta_{\min}.
\end{align}
We compute the solution numerically by the Broyden's method \citep{broyden1965class} implemented in the \proglang{R} package \pkg{nleqslv} \cite{hasselman1nleqslv}. Let $s_0$ be the solution of (\ref{eq:saddle-eq}), $\lambda_3=\kappa^{(3)}_{\delta}(s_0)/(\kappa_{\delta}^{(2)}(s_0))^{3/2}$, $W=\textrm{sign}(s_0) \sqrt{2(\delta_{\min}s_0-\kappa_\delta(s_0))}$ and $Q=s_0\sqrt{\kappa^{(2)}_{\delta}(s_0)}$,  then \cite{huang2011saddlepoint} suggests different approximations based on the sign of $s_0$. If $s_0$ is zero, then $saEI(\bx) = \sqrt{\kappa^{(2)}_{\delta}(0) / 2\pi}$, whereas, if $s_0$ is positive, then
\begin{eqnarray}\label{eq:saddle-approx-pos}
saEI(\bx)&\approx &
  \delta_{\min}-\mu_{\delta} + e^{-\frac{W^2}{2}}\left\{\sqrt{\dfrac{\kappa_{\delta}^{(2)}(s_0)}{2\pi}}-s_0 \kappa_{\delta}^{(2)}(s_0)\ e^{\frac{Q^2}{2}}[1-\Phi(Q)]\right\} \\ \nonumber
   && +\ e^{\frac{Q^2-W^2}{2}}\sqrt{\kappa_{\delta}^{(2)}(s_0)}\ \dfrac{\lambda_3}{6}\left\{[1-\Phi(Q)](Q^4+3Q^2)-\phi(Q)(Q^3+2Q)\right\},
\end{eqnarray}
and for negative $s_0$, 
\begin{eqnarray}\label{eq:saddle-approx-neg}
saEI(\bx)&\approx &
     e^{-\frac{W^2}{2}}\left\{\sqrt{\frac{\kappa_{\delta}^{(2)}(s_0)}{2\pi}}+s_0\kappa^{(2)}_{\delta}(s_0)e^{\frac{Q^2}{2}}\Phi(Q)\right\} \\ \nonumber
    && -\ e^{\frac{Q^2-W^2}{2}}\sqrt{\kappa^{(2)}_{\delta}(s_0)}\ \dfrac{\lambda_3}{6}\left\{\Phi(Q)(Q^4+3Q^2)+\phi(Q)(Q^3+2Q)\right\}
\end{eqnarray}
where $\Phi(\cdot)$ and $\phi(\cdot)$ are the cumulative distribution function and probability density function of the standard normal distribution (see \cite{huang2011saddlepoint} for details). The built-in function called \code{saEI} in the \proglang{R} package \pkg{DynamicGP} facilitates easy computation of this criterion for choosing follow-up design points \citep{dynamicGP}.

%It is straightforward to verify that, for all $s$ such that $(\hat{\bSig}^{-1}-2s\bI_L)$ is positive definite, $\kappa_{\delta}^{(2)}(s)>0$, and  $\kappa^{(1)}_{\delta}(s)$ range from zero to infinity over the prescribed span of $s$. Thus, the saddlepoint equation (\ref{eq:saddle-eq}) has an unique solution given $K_{\min}>0$. {\textcolor{red}{why?}

%\textcolor{red}{Unlike the EI expressions in \cite{jones1998efficient} or \cite{ranjan2008sequential}, the tradeoff between exploration and exploitation can not be easily established using different terms of $saEI(\bx)$. }

\textbf{Remark:} As a special case, if the SVD of $\bY$ has only one non-zero singular value, the EI criterion has a closed form. From an application standpoint, a simulator with a separable structure, $f(\bx,t)=f_1(\bx)f_2(t)$, leads to exactly one non-zero singular value, provided both $f_1$ and $f_2$ are non-zero functions.  Result~\ref{result:ei-trivial} presents the exact EI expression (\ref{eq:ei-crit}) in such a scenario.

\begin{result}\label{result:ei-trivial}
  Suppose there is only one non-zero singular value $d_1$ in the SVD of the
  response matrix $\bY$ of the dynamic computer simulator, then the
  EI criterion (\ref{eq:ei-crit}) simplifies to
\begin{align}\label{eq:trivial-ei}
  \begin{aligned}
  E[I(\bx)]=&\left[\delta_{\min}-\bxi^T\bxi+2\bxi^T\bb_1\hat{c}_1(\bx)-d_1^2\hat{c}^2_1(\bx)-d_1^2\hat{\sigma}^2_1(\bx)\right]\left[\Phi(l_2(\bx))-\Phi(l_1(\bx))\right]\\
  &+2\left[d_1^2\hat{c}_1(\bx)\hat{\sigma}_1(\bx)-\bxi^T\bb_1\hat{\sigma}_1(\bx)\right]\left[\phi(l_2(\bx))-\phi(l_1(\bx))\right]\\
  &+d_1^2\hat{\sigma}_1^2(\bx)\left[l_2(\bx)\phi(l_2(\bx))-l_1(\bx)\phi(l_1(\bx))\right],
  \end{aligned}
\end{align}
when $(\bxi^T\bb_1)^2+d_1^2(\delta_{\min}-\bxi^T\bxi)>0$ and $E[I(\bx)]=0$
otherwise, where $l_1(\bx)=(w_1-\hat{c}_1(\bx))/\hat{\sigma}_1(\bx)$, $l_2(\bx)=(w_2-\hat{c}_1(\bx))/\hat{\sigma}_1(\bx)$, and
{\color{black}
\begin{align*}
  \begin{aligned}
  &w_1=\frac{\bxi^T\bb_1-\sqrt{(\bxi^T\bb_1)^2+d_1^2(\delta_{\min}-\bxi^T\bxi)}}{d_1^2}, \
  w_2=\frac{\bxi^T\bb_1+\sqrt{(\bxi^T\bb_1)^2+d_1^2(\delta_{\min}-\bxi^T\bxi)}}{d_1^2}.
  \end{aligned}
\end{align*}
}
\end{result}
  
Appendix~B presents the derivation of the EI criterion in (\ref{eq:trivial-ei}). It is important to note that the separable  structure of the simulator is not very realistic and one may have to rely on the saddlepoint approximation based EI (saEI) criterion.

\subsection{Extraction of the Inverse Solution}
\label{sec:sol-extract}

At the end of the sequential design procedure, the final fit with the MAP estimates of the model parameters $\bm{\Theta}$, $\sigma^2_i$'s  and $\sigma^2$ are used to extract the best estimate of the inverse solution, i.e., the estimated optimal input that corresponds to the model output with the best match/approximation of the target response $\bxi$.

A naive approach is to use $\tilde{\bx}^*=\underset{\bx\in\Omega}{\mathrm{argmin}}\:\|\bxi-\hat{\by}(\bx)\|^2_2\ $ as the best estimate of the inverse solution, where
$\hat{\by}(\bx)=\tE[\by(\bx)|\bY]=\bB\hat{\bc}(\bx|\bV^*,\hbT)$ is the mean prediction in (\ref{eq:final}) based on $N=n_0 + n_{new}$ points. 
Let $\hat{\bc}_{\bxi}=\bD^{*-2}\bB^T\bxi$ and $\hat{c}_{\bxi,i}$ be the $i$th entry of $\hat{\bc}_{\bxi}$.  Then, one can easily show that
\begin{align}\label{eq:naive}
  \begin{aligned}
  \|\bxi-\hat{\by}(\bx)\|^2_2&= \|\bxi-\bB\hat{\bc}_{\bxi} + \bB\hat{\bc}_{\bxi} - \bB\hat{\bc}(\bx|\bV^*,\hbT)\|_2^2\\
  &=\|\bxi-\bB\hat{\bc}_{\bxi}\|_2^2 + \|\bB\hat{\bc}_{\bxi}-\bB\hat{\bc}(\bx|\bV^*,\hbT)\|_2^2 \\
  &=\|\bxi-\bB\hat{\bc}_{\bxi}\|_2^2 + \sum_{i=1}^pd_i^2\big(\hat{c}_i(\bx|\bv_i,\hbt_i)-\hat{c}_{\bxi,i}\big)^2.
  \end{aligned}
\end{align}
Since $\|\bxi-\bB\hat{\bc}_{\bxi}\|_2^2$ is a constant with respect to $\bx$, the optimal inverse solution is 
$$\tilde{\bx}^*=\underset{\bx\in\Omega}{\mathrm{argmin}}\:\sum_{i=1}^pd_i^2\big(\hat{c}_i(\bx|\bv_i,\hbt_i)-\hat{c}_{\bxi,i}\big)^2.$$

The function called \code{SL2D} (squared $L_2$ discrepancy) in \proglang{R} package \pkg{DynamicGP} provides the implementation of this naive approach for the SVD-based GP model.

The naive approach, however, ignores the presence of the predictive uncertainty due to surrogate modelling. Thus we propose minimizing the expected squared $L_2$ discrepancy (ESL2D) between the field (or target) observation and the dynamic computer simulator outputs, i.e.,
\begin{align*} 
  \hat{\bx}^* = \underset{\bx\in\Omega}{\mathrm{argmin}}\: \tE\big[\delta(\bx)\big|\bY\big],
\end{align*}
where the expectation is taken with respect to the predictive distribution (\ref{eq:final}). This is precisely $\mu_{\delta}$ in Result~1. Focussing only on the terms that depend on $\bx$, the inverse solution is given by
\begin{align}\label{eq:inv-ESL2D}
  \hat{\bx}^* =  \underset{\bx\in\Omega}{\mathrm{argmin}}\: 
    \left[\sum_{i=1}^pd_i^2 (\hat{c}_i(\bx|\bv_i,\hbt_i)-\hat{c}_{\bxi,i})^2 + \sum_{i=1}^pd_i^2 \hat{\sigma}^2_i(\bx|\bv_i,\hbt_i) \right].
\end{align}

 It is important to note that the first term is same as the naive criterion and the second term represents the total weighted prediction uncertainty at $\bx$. The function \code{ESL2D} in the \proglang{R} package \pkg{DynamicGP} implements this extraction method.  An empirical study based on a few test functions shows that the proposed extraction method via ESL2D is more accurate than the naive approach  in extracting the best estimate of the inverse solution. The simulation results are presented in the Supplementary Material.

As the run size approaches to infinity, it is sensible to conclude that both solutions by the naive approach and the proposed ESL2D approach can converge to the true optimal solution.  Theorem~\ref{thm-chedan} establishes the convergence of the ESL2D criterion to the true inverse solution.

\begin{theorem}\label{thm-chedan}
Let $\by(\bx_i)$ be the time-series valued simulator response for $\bx_i \in \Omega=[0,1]^q$, for $i=1,...,N$, and $\bxi$ be the pre-specified target output. If $\by(\bx)$ is emulated via the SVD-based GP model in Section~\ref{sec:svd-gpm}, and the discrepancy $\delta(\bx)$ is a scalar valued continuous process, then the ESL2D optimal solution $\hat{\bx}^*$ in (\ref{eq:inv-ESL2D}) converges in probability to the true solution $\bx^*$ in (\ref{eq:ls-sol}), as $N \rightarrow \infty$.
\end{theorem}

See Appendix~C for the proof of Theorem~\ref{thm-chedan}. In practice,  the inverse solution in (\ref{eq:inv-ESL2D}) can either be extracted by the exhaustive evaluation of \code{ESL2D} on a large candidate set in the design domain as suggested by \citet{pratola2013fast} and \citet{sinsbeck2017sequential}, or by a global optimization algorithm such as the genetic algorithm, particle swarm optimization method, or the branch and bound algorithm \citep{banzhaf1998genetic, franey2011branch}.

\section{Simulated Examples and TDB Application}
\label{sec:sim-study}

In this section, we use \nsimexp test functions and the TDB application based computer simulator for a thorough comparison between the proposed sequential approach (denoted as saEI) and the popular competitors for solving the inverse problem. For a fair comparison, the competitors considered here are also sequential in nature. 

\cite{ranjan2016inverse} suggested treating the squared $L_2$ discrepancy $\delta(\bx)=\|\bxi-\by(\bx)\|_2^2$, for each $\bx$ in the design space, as the scalarized simulator output, and then found the minimizer of $\delta(\bx)$ via GP modeling coupled with the EI criterion for global optimization \citep{jones1998efficient, picheny2013quantile, ranjan2013}. Alternatively, \citet{pratola2013fast} worked with the logarithm of the ratio of two likelihoods, one where $\bx$ corresponds to $\bxi - \by(\bx) = \be$ (i.e., the desired inverse solution), and the other one where $\bxi - \by(\bx) \ne \be$, with $\be$ being the random noise. This scalarized log-likelihood ratio statistic was emulated via GP model and then minimized using the EI criterion in \cite{jones1998efficient}. In this paper, we denote these two approaches by SL2 and LR methods, respectively.

For simulated examples, the performance comparison can be measured by the normalized discrepancy between the simulator response at the estimated inverse solution, $\by(\hat{\bx}^*)$, and the target response, $\bxi$, defined as:
\begin{align*}
D_{\bxi}=\frac{\|\bm{\xi}-\by(\hat{\bx}^*)\|^2_2}{\|\bm{\xi}-\bar{\xi}\bm{1}_L\|^2_2},
\end{align*}
where $\bar{\xi}=\sum_{i=1}^L\xi_{t_i}/L$, and $\bm{1}_L$ is an
$L$-dimensional vector of ones. %\textcolor{red}{What have we done for the real application?}

The results presented in this section are averaged over 50 simulations. For each simulated example, we randomly choose an $\bx^*$ in the design domain $\Omega\subseteq\mathbb{R}^q$, and keep it fixed throughout the 50 simulations. For implementation purpose, we rescale the inputs to $[0, 1]^q$. The setup of the simulation study ensures the numbers of simulator evaluations for the three methods to be exactly the same for a fair comparison. Based on some preliminary exploration of the test functions considered here, we have specified $n_0=6q$ and $n_{new}=N-n_0=12q$ as the size of initial designs and the number of follow-up points, respectively. The initial designs are selected using a maximin Latin hypercube design (LHD) generated by the R package $lhs$ \citep{lhs}.  All the three methods use the same initial design in each simulation. The remaining $n_{new}$ design points are added sequentially one at-a-time as per their respective criteria.

The field observations (i.e., the targets) vary slightly across the 50 simulations, as the error terms in $\bxi=\by(\bx^*)+\bm{e}$, are generated randomly, $\bm{e}\sim\mathcal{N}(\bm{0},\rho\Var(\by(\bx^*))\bI_L)$, where $1/\rho$ is the signal to noise ratio - set to 50 for all simulated examples. For sequential selection of the follow-up points, we use a large candidate set of size $M_1 = 2000q$ in the design domain chosen randomly for every simulation.
The final estimated inverse solutions are extracted using a different candidate set of size $M_2 = 2000q$, which is also randomly chosen over different simulations.

\subsection{Example 1: One-dimensional  Example}
\label{sec:illu-ex-cont}

Suppose that the simulator outputs are obtained using
\begin{align}\label{eq:illu-examp}
  y_t(x)=\frac{\sin((8x+6)\pi t)}{2t}+(t-1)^4,
\end{align}
where $x\in[0,1]$ and $t\in[0.5,2.5]$ is on a 200-point equidistant time grid. The simulator response at an input $x$ is $\by(x)=[y_{t_1}(x),\dots,y_{t_{200}}(x)]^T$. The field observation $\bxi=[\xi_{t_1},\dots,\xi_{t_{200}}]^T$ is generated via $\xi_{t_i}=y_{t_i}(x^*)+e_i$, where $x^*=0.7861$ is randomly chosen using a uniform distribution on [0,1], and the observational error terms $e_i$'s are Gaussian white noise with mean 0 and variance $\sigma^2_{e}=0.03363$, which produces a signal to noise ratio 50 $(=\sigma^2_{\by(x^*)}/\sigma^2_{e}$), with $\sigma^2_{\by(x^*)} = \Var(\by(x^*))=\sum_{i=1}^L(y_{t_i}(x^*)-\bar{y}(x^*))^2/(L-1)$ and $\bar{y}(x^*)=\sum_{i=1}^Ly_{t_i}(x^*)/L$. The sequential design approach starts with fitting the SVD-based GP surrogate to a 6-point maximin LHD and selects $N-n_0=12$ follow-up points by maximizing the proposed saEI criterion.

%  \textcolor{red}{say something about saddlepoint approximation?}

\begin{figure}[h!]
  \centering
  \includegraphics[width=6.5in]{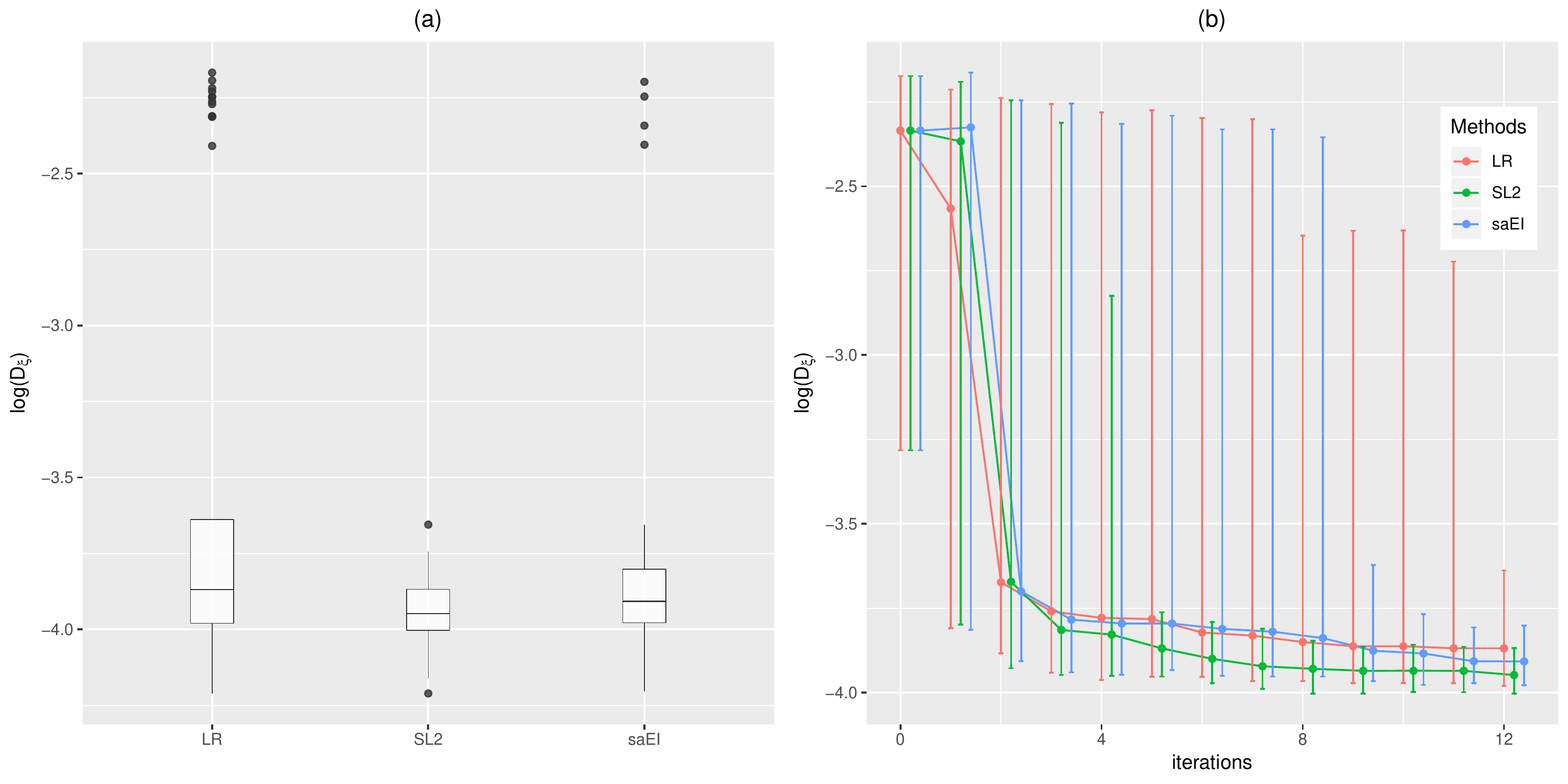}
  \caption{Example 1: (a)  the boxplot of  the $\log(D_{\bxi})$  values from the final fitted surface, and (b) the median, first and third quartile of the $\log(D_{\bxi})$ values over 50 repetitions obtained from the fitted surface in different iterations of the sequential design procedure.}
\label{fig:illu-cont}
\end{figure}

 Figure~\ref{fig:illu-cont} summarizes the  $\log(D_{\bxi})$ values over 50 simulations for the three methods (LR, SL2 and saEI). Figure~\ref{fig:illu-cont}(a) shows the boxplot of the $\log(D_{\bxi})$ values obtained from the final estimate of the inverse solution at the end of the sequential procedure. Figure~\ref{fig:illu-cont}(b) presents the three quartiles of the $\log(D_{\bxi})$  values (over 50 simulations) after every follow-up point has been selected in the sequential design procedure. Figure~\ref{fig:illu-cont}  indicates the SL2 and saEI methods provide similar performance which is slightly better than the LR method in terms of the median accuracy of the solution of the inverse problem as well as the convergence of the solution.  It is also interesting to note from Figure~\ref{fig:illu-cont}(b) that the SL2 method  appears to be more stable than the other two methods.

\subsection{Example 2 \citep{harari2014convex}}

\begin{figure}[h!]
  \centering
  \includegraphics[width=6.5in]{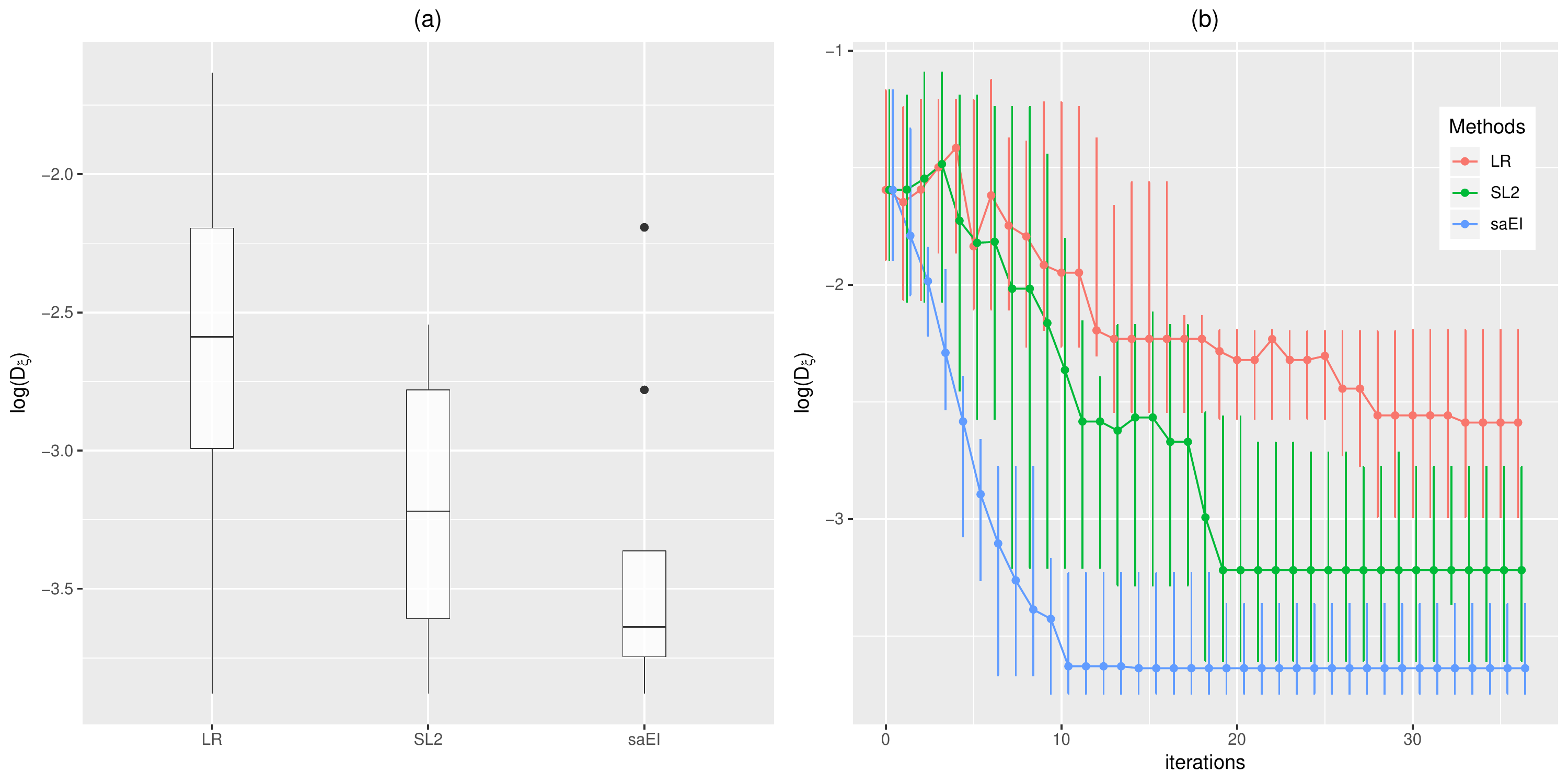}
  \caption{Example~2: (a) the boxplots of the $\log(D_{\bxi})$ values from the final fitted surface, and (b) the median, first and third quartile of the $\log(D_{\bxi})$ values over 50 repetitions obtained from the fitted surface in different iterations of the sequential design procedure.}
\label{fig:hardev}
\end{figure}

Suppose the outputs of the dynamic computer simulator obey
\begin{align}\label{eq:harari}
  y_t(\bx) = \exp(3x_1t+t)\cos(6x_2t+2t-8x_3-6),
\end{align}
where $\bx=(x_1,x_2,x_3)^T\in[0,1]^3$ and $t\in[0,1]$ is on a
200-point equidistant time-grid. The input producing the target (or, equivalently, the field
observation) is randomly generated as $\bx^*=[0.522,0.950,0.427]^T$. Here, the initial design is of size $n_0=6\times3 = 18$, and $N-n_0= 36$ points were chosen sequentially one at-a-time as per the individual design criterion (e.g., using $saEI(\bx)$).
Figure~\ref{fig:hardev} summarizes the simulation results over 50 replications. %A different initial design, error term in the target response, candidate set for the choice of follow-up design and the final candidate set for extracting the inverse solution are chosen randomly in each realization.

The two panels of Figure~\ref{fig:hardev} present the performance comparison of the three methods in terms of the boxplot of the normalized discrepancy $\log(D_{\bxi})$ from the final estimate of the inverse solution and the intermediate estimates from different iterations of the sequential design procedure.
Figure~\ref{fig:hardev}(a) shows that the proposed approach (saEI) produces significantly smaller $\log(D_{\bxi})$ values as compared to the other methods. Panel (b) indicates that the proposed approach is not only giving the best final inverse solution estimate, but also it achieves higher accuracy at a faster rate, and the solution is reliable (with less variability). It is also interesting to note from Figure~\ref{fig:hardev}(a) that the worst 25-th percentile of the solution found by the proposed method is slightly better than the best 25-th percentile solution obtained using the  LR approach, and the SL2 technique can sometimes give very good estimates as compared to the proposed saEI method.

\subsection{Example 3 \citep{bliznyuk2008bayesian}}

\begin{figure}[h!]
  \centering
  \includegraphics[width=6.5in]{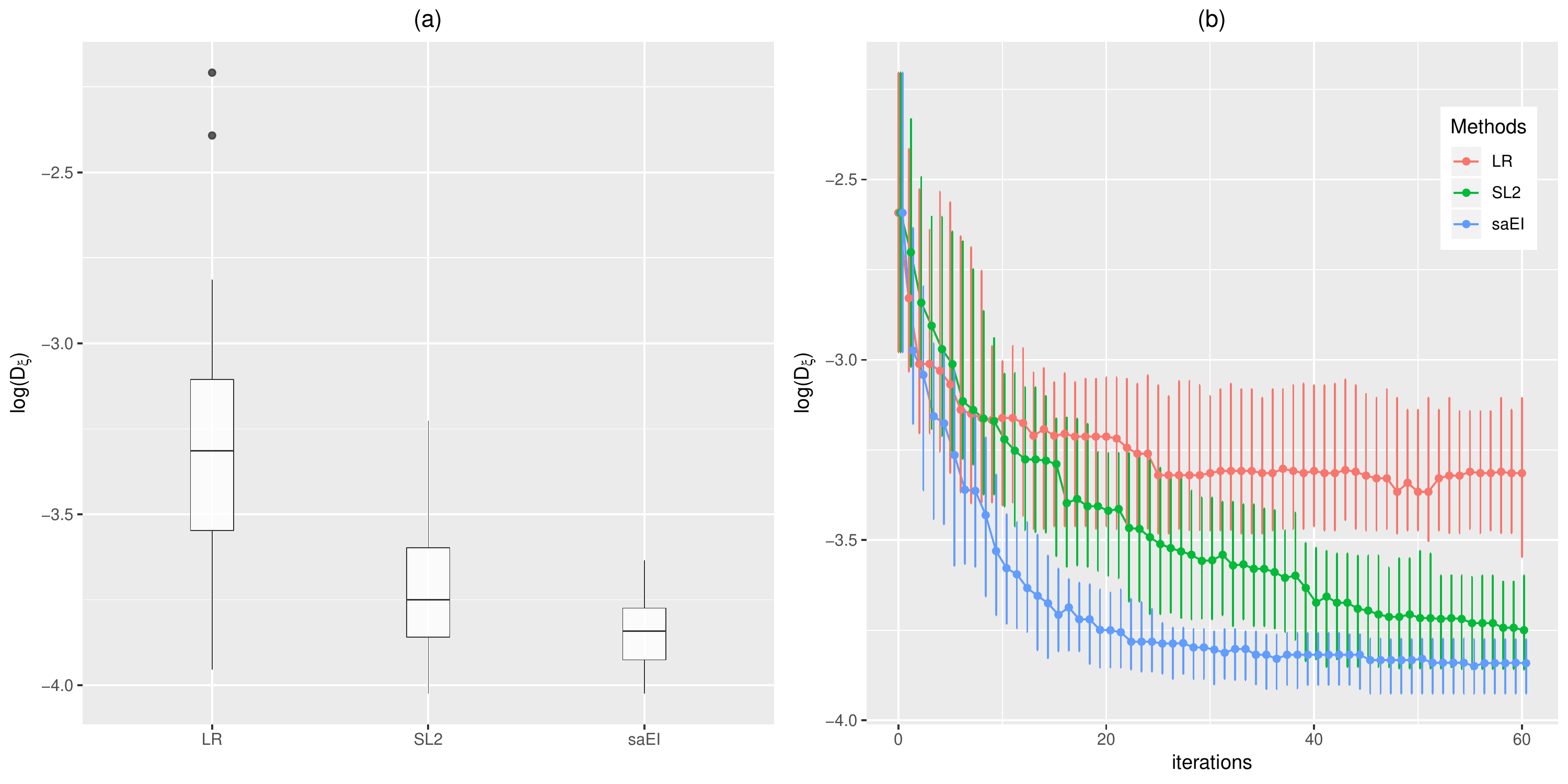}
  \caption{Example 3: (a) the boxplots  of the $\log(D_{\bxi})$ values from the final fitted surface, and (b) the median, first and third quartile of the $\log(D_{\bxi})$ values over 50 repetitions obtained from the fitted surface in different iterations of the sequential design procedure.}
\label{fig:envdev}
\end{figure}

Consider the environmental model by \cite{bliznyuk2008bayesian} which simulates a pollutant spill at two locations ($0$ and $L$) caused by a chemical accident. The computer simulator outputs are generated using the following model that captures concentration at space-time point $(s,t)$,
\begin{align}\label{eq:environ}
 \begin{aligned}
  y_t(\bx)&=C(s,t; M,D,L,\tau) \\
&= \frac{M}{\sqrt{Dt}}\exp\left(\frac{-s^2}{4Dt}\right)+\frac{M}{\sqrt{D(t-\tau)}}\exp\left(-\frac{(s-L)^2}{4D(t-\tau)}\right)I(\tau<t),
 \end{aligned}
\end{align}
where $\bx=(M,D,L,\tau,s)^T$, $M$ denotes the mass of pollutant
spilled at each location, $D$ is the diffusion rate in the chemical
channel, and ($0$ and $\tau$) are the time of the two spills. The input domain is
$\bx\in[7,13]\times[0.02,0.12]\times[0.01,3]\times[30.01,30.295]\times[0,3]$, and
$t\in[0.3,60]$ lies on a regular 200-point equidistant time-grid. In
this example, the randomly chosen input that produces the field observation is $\bx^*=[9.676,0.05947,1.456,30.27,2.532]^T$. Following the general rule outlined earlier, we used a 30-point initial design and $60$ sequential points to estimate the inverse solution.
Figure~\ref{fig:envdev} summarizes the results of 50 simulations.

Figure~\ref{fig:envdev}(a) shows that the proposed saEI approach clearly outperforms the two competitors. As in Example~2, the  $\log(D_{\bxi})$ values show significantly faster convergence for the proposed method, and the trend in the relative performance is somewhat consistent.

\subsection{TDB Application}\label{sec:app-tdb}

In this section, we investigate the performance of the proposed approach for calibrating the TDB model with respect to the target response on average juvenile count over a season spanning a period of 102 Julian days. After extensive deliberation and screening of the input variables, the following six input variables  have been identified for the calibration of the TDB model:
\begin{itemize}
\item $\mu_4$ -- adult death rate,
\item $\beta$ -- maximum fecundity (eggs laid per day),
\item $\nu$ -- non-linear crowding parameter,
\item $\tau_1$ -- first delay - hatching time of summer eggs,
\item $\tau_2$ -- second delay - time to maturation of recently
  hatched eggs,
\item Season -- average number of days on which adults switch
  to laying winter eggs.
\end{itemize}

\begin{figure}[h!]
  \centering
  \includegraphics[width=6.5in]{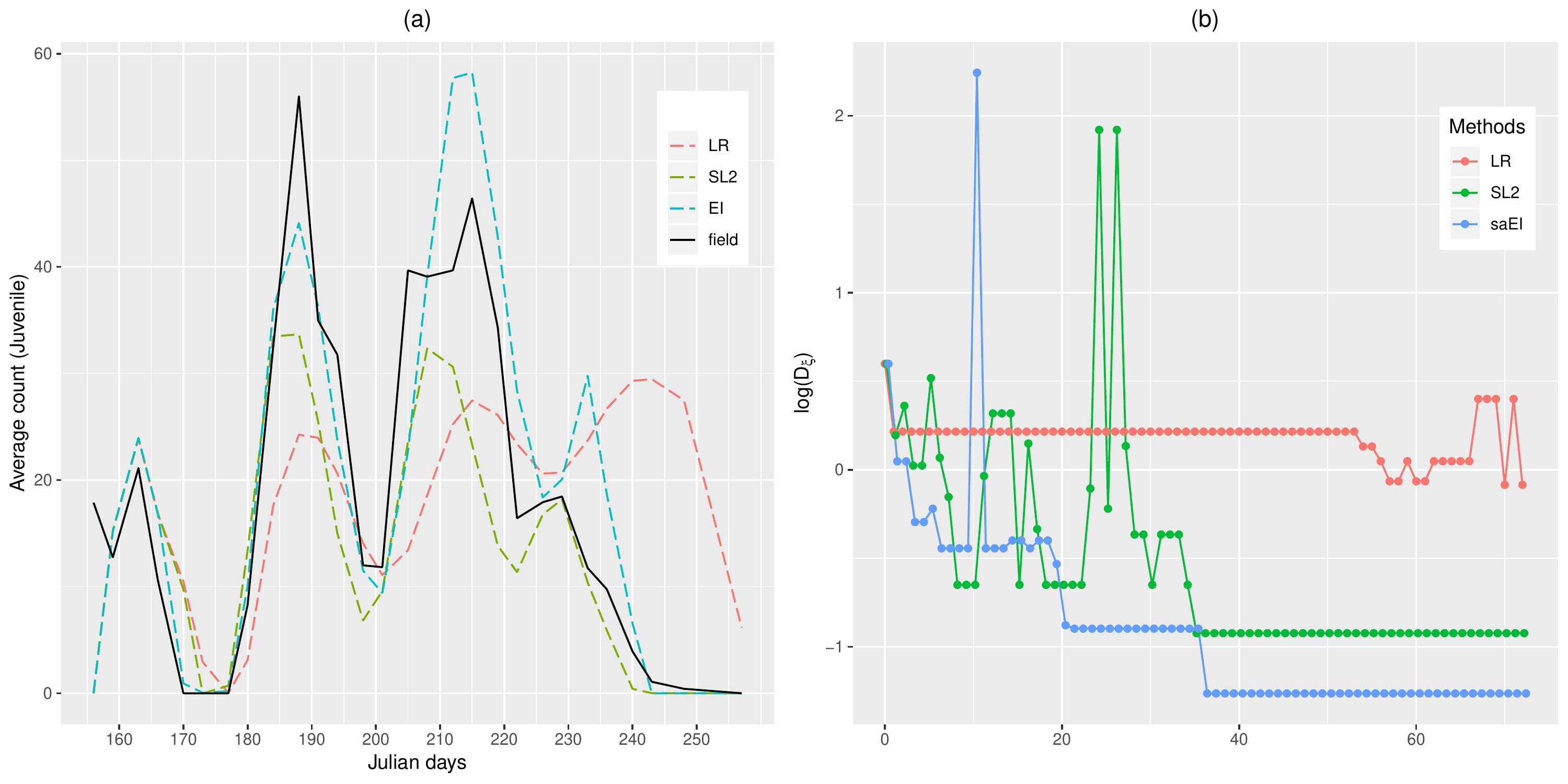}
  \caption{(a) The field observation of the juvenile ERM population
    evolution and the located simulator outputs that match the field
    observation produced by the three sequential design methods.  (b)
    The traces of the $\log(D_{\bxi})$ criterion values of  the 72 iterations for the three methods.}
\label{fig:tdb}
\end{figure}

In our study, the three methods (the proposed saEI, and the competitors SL2 and LR) start from the same initial design, a maximin LHD with $n_0=6q=36$ points, and sequentially add $12q=72$ points one at-a-time. The simulator outputs at the estimated inverse solutions produced by the three methods are shown in panel (a) of Figure~\ref{fig:tdb}. Panel
(b) of the figure provides the traces of $\log(D_{\bxi})$ values for each of the 72 iterations.

As shown in panel (a), the proposed method produces significantly better match for the field observation at the hump between 200 and 220 Julian days compared with the other methods.  Panel (b) shows that the proposed method produces the smallest minimum $\log(D_{\bxi})$ and leads to faster convergence towards the minimum discrepancy $\delta(\bx)$.
  
\section{Concluding Remarks}
\label{sec:con-rmk}

In this article, we have proposed an effective sequential design approach for the inverse problem of dynamic computer experiments which selects follow-up design points by maximizing the proposed saEI criterion, and extracts the inverse solution by minimizing a novel ESL2D criterion from the final SVD-based GP surrogate. The asymptotic property of the proposed inverse solution is estimated. The proposed method is shown empirically to be more accurate in estimating the solution of the inverse problem as compared to the two existing sequential design approaches, LR \citep{pratola2013fast} and SL2 \citep{ranjan2016inverse}. Moreover, the proposed saEI criterion is computationally efficient to evaluate by the saddlepoint approximation technique.

There are a few issues worth addressing. First, the history matching
(HM) algorithm \citep{vernon2010galaxy,bhattacharjee2017} is not
included into our comparison for the following two main reasons. One 
is, the HM algorithm is essentially not a standard sequential design
approach. Instead of sequentially including points to the design set,
it sequentially eliminates implausible points from a large candidate
set. The other reason is, the HM algorithm does not necessarily
provide a single point as an estimate of the solution of the inverse
problem. It provides a set of non-implausible points which might be
empty or consist of multiple points.

Second, in each of the four examples studied in Section
\ref{sec:sim-study}, the three sequential design methods select the
same fixed number of the follow-up design points. This setup is
applied out of consideration for the fairness of comparison. In
some applications, it might be more desirable to terminate the iteration of
the sequential design when it is determined to be convergent by some
stopping criteria. However, we have not discovered a simple stopping
criterion that works for all the three methods in the four
examples. To the best of our knowledge, the choice of stopping
criteria is on an ad hoc basis which requires expert knowledge on the
specific problem.

Third, due to the growth of the computational power, the dynamic
computer simulator could be evaluated at a large input data set ($N$
is large). The big $N$ issue prohibits the fitting of the SVD-based GP
model since the computational cost of the empirical Bayesian inference
is $O(N^3)$ computational complexity. To address this issue,
\cite{zhang2017local} proposes a local approximate SVD-based GP model. Its application for the inverse problem is a work in progress.

Fourth, the proposed approach imposes a restriction that the field
observation and computer simulator outputs are collected at the same
set of time points. In some real-life applications, the two sources of
data may not always align. Such case can occur because the field
observation at certain time points might be missing, or, the computer
model outputs may not be sampled on regular grids, which means the
outputs of the computer simulator are not observed on the same set of
time points \citep{hung2015analysis}. For the case of missing time
points in the field observation, the expectation maximization (EM)
algorithm \citep{dempster1977maximum} can be applied for evaluation of
the projection coefficient $\hat{\bc}_{\bxi}$ of the field observation. When the
outputs of the computer model are not sampled on regular grids,
\cite{hung2015analysis} provides a feasible method based on
spatio-temporal GP models with separable correlation functions. An innovative generalization of the proposed methodology in this scenario is a possible avenue for
our future research.

\appendix

\renewcommand{\thesection}{Appendix \Alph{section}}
\renewcommand{\theequation}{A.\arabic{equation}}
\renewcommand{\theproposition}{A.\arabic{proposition}}

\setcounter{section}{0}
\setcounter{equation}{0}
\setcounter{result}{0}
\setcounter{proposition}{0}
\section{Moments of Discrepancy Function}
\label{sec:moments_cgf}

Here we present the derivation of Results~1 and 2 (in Section~4.1). Result~1 presents the expression for expected squared $L_2$ discrepancy with respect to the SVD-based GP surrogate (in Section~3) and Result~2 shows the closed form expression for the cumulant generating function of $\delta(\bx)$. 

%
%\begin{result}\label{result:mu_delta}
%The expected discrepancy with respect to the predictive distribution (\ref{eq:final})  is given by 
%% 
%\begin{align*}
%  \mu_\delta &= \tE[\delta(\bx)|\bY]  =\bxi^T (\bI_L - \bU^*\bU^{*T})\bxi  + \sum_{i=1}%^pd_i^2\left[(\hat{c}_i(\bx)-\hat{c}_{\bxi,i})^2 + \hat{\sigma}^2_i(\bx)\right] +\hat{\sigma}%^2L. 
%\end{align*}
%%
%where $\hat{\bc}_{\bxi}=\bD^{*-2}\bB^T\bxi$.
%\end{result}
%------------------------

\begin{proof}[Proof of Result~1]
For an arbitrary $\bx\in \Omega$, since $\delta(\bx) = \|\bxi - \by(\bx)\|^2_2 $ is a quadratic form and the predictive distribution of $\by(\bx)$ given $\bY$ is 
$$\pi(\by(\bx)|\bY) \approx \mathcal{N}\big(\bm{B}\hat{\bm{c}}(\bm{x}_0|\bV^*,\hat{\bm{\Theta}}),\bB\bm{\Lambda}(\bV^*,\hat{\bm{\Theta}})\bB^T+\hat{\sigma}^2\bI_L\big),$$
the corresponding expected squared $L_2$ discrepancy is 
\begin{align*}
  \mu_\delta &= \tE[\delta(\bx)|\bY]  = \tE[(\bxi-\by(\bx))^T(\bxi-\by(\bx))|\bY] \nonumber \\
                      &=(\tE[\bxi-\by(\bx)|\bY])^T(\tE[\bxi-\by(\bx)|\bY])+\ttr(\bB\bLam(\bV^*,\hbT)\bB^T+\hat{\sigma}^2\bI_L)\nonumber  \\
                      &=\left(\bxi-\bB\hat{\bc}(\bx)\right)^T\left(\bxi-\bB\hat{\bc}(\bx)\right)+ \ttr\left(\bLam(\bV^*,\hbT)\bB^T\bB\right)+\hat{\sigma}^2L.  
\end{align*}
Orthogonality of the columns of $\bU^*$ implies that $\|\bxi-\bB\hat{\bc}\|_2^2= \|\bxi-\bB\hat{\bc}_{\bxi}\|_2^2 + \|\bB\hat{\bc}_{\bxi}-\bB\hat{\bc}(\bx)\|_2^2 $, and $\|\bxi-\bB\hat{\bc}_{\bxi}\|_2^2
= \bxi^T\bxi - \bxi^T\bU^*\bU^{*T}\bxi$. Thus,
\begin{align*}           %
  \mu_\delta&=\|\bxi-\bB\hat{\bc}_{\bxi}\|_2^2 + \|\bB\hat{\bc}_{\bxi}-\bB\hat{\bc}(\bx)\|_2^2  + \ttr\left(\bLam(\bV^*,\hbT)\bB^T\bB\right)+\hat{\sigma}^2L \nonumber  \\ 
         &=\|\bxi-\bB\hat{\bc}_{\bxi}\|_2^2 +\sum_{i=1}^pd_i^2\big(\hat{c}_i(\bx)-\hat{c}_{\bxi,i}\big)^2 +\sum_{i=1}^pd_i^2\hat{\sigma}^2_i(\bx)+\hat{\sigma}^2L \\
         &=\bxi^T (\bI_L - \bU^*\bU^{*T})\bxi +\sum_{i=1}^pd_i^2\left[ \big(\hat{c}_i(\bx)-\hat{c}_{\bxi,i}\big)^2 + \hat{\sigma}^2_i(\bx)\right]+\hat{\sigma}^2L. 
\end{align*}
\end{proof} 

%===================================
%\begin{result} 
%The cumulant generating function of $\delta(\bx)=\|\bxi-\by(\bx)\|^2_2$ with respect to the svd-based GP surrogate in (\ref{eq:final}) can be simplified to 
%%
%\begin{align*}
%  \kappa_{\delta}(s)=&-\frac{L-p}{2}\log(1-2s\hat{\sigma}^2)-\frac{1}{2}\sum_{i=1}^p
%\log(1-2s(\hat{\sigma}^2+\hat{\sigma}^2_id_i^2))\\ \nonumber
%  %
%  &+\sum_{i=1}^p\frac{2s^2\hat{\sigma}_i^2 \mu^2_{bi}}{[1-2s(\hat{\sigma}_i^2d_i^2+
%\hat{\sigma}^2)](1-2s\hat{\sigma}^2)} + \frac{s\bmu^T\bmu}{1-2s\hat{\sigma}^2},
%\end{align*}
%where $\mu_{bi}$ is the $i$-th entry of $\bB^T\bmu$.
%%
%\end{result}
%------------------------
\begin{proof}[Proof of Result~2]
Recall that if $\hat{\bSig}^{-1}-2s\bI_L$ is positive definite, Lemma B.1.1 of \cite{santner2003design} gives the moment generating function of $\delta(\bx) = (\bxi-\by(\bx))^T(\bxi-\by(\bx))$ as
\begin{align*} 
  M_{\delta}(s)=|\bI_L-2s\hat{\bSig}|^{-\frac{1}{2}}\exp\left\{\frac{\bmu^T\hat{\bSig}^{-1}(\hat{\bSig}^{-1}-2s\bI_L)^{-1}\hat{\bSig}^{-1}\bmu-\bmu^T\hat{\bSig}^{-1}\bmu}{2}\right\},
\end{align*}
which further leads to the cumulant generating function of $\delta(\bx)$ given by
\begin{align*}
  \kappa_{\delta}(s)={-\frac{1}{2}}\log(|\bI_L-2s\hat{\bSig}|) + \frac{1}{2}\ \bmu^T\hat{\bSig}^{-1}(\hat{\bSig}^{-1}-2s\bI_L)^{-1}\hat{\bSig}^{-1}\bmu - \frac{1}{2}\ \bmu^T\hat{\bSig}^{-1}\bmu .
\end{align*}

In the first term of $\kappa_{\delta}(s)$, we have 
\begin{align*}
  |\bI_L-2s\hat{\bSig}|&=|(1-2s\hat{\sigma}^2)\bI_L-2s\bB\bm{\Lambda}\bB^T|
               =(1-2s\hat{\sigma}^2)^L|\bI_L-\frac{2s}{1-2s\hat{\sigma}^2}\bB\bLam\bB^T|\\
               &=(1-2s\hat{\sigma}^2)^L|\bI_p-\frac{2s}{1-2s\hat{\sigma}^2}\bB^T\bB\bLam|\\
&\textrm{(from Sylvester's determinant identity, Lemma 2.8.6 of \cite{bernstein2005matrix})}             \\  
               &=(1-2s\hat{\sigma}^2)^L|\bI_p-\frac{2s}{1-2s\hat{\sigma}^2}\bD^{*2}\bLam|
               =(1-2s\hat{\sigma}^2)^{L-p}\prod_{i=1}^p\left(1-2s(\hat{\sigma}^2+\hat{\sigma}^2_id_i^2)\right).
\end{align*}

The central part of the second term, $\bmu^T\hat{\bSig}^{-1}(\hat{\bSig}^{-1}-2s\bI_L)^{-1}\hat{\bSig}^{-1}\bmu$, is
\begin{align*}
\hat{\bSig}^{-1}(\hat{\bSig}^{-1}-2s\bI_L)^{-1} &= (\bI_L-2s\hat{\bSig})^{-1}
  =\left((1-2s\hat{\sigma}^2)\bI_L-2s\bB\bm{\Lambda}\bB^T\right)^{-1}\\
  &=(1-2s\hat{\sigma}^2)^{-1}\bI_L-(1-2s\hat{\sigma}^2)^{-1}\bB\left(\bD^{*2}-\frac{1-2s\hat{\sigma}^2}{2s}\bLam^{-1}\right)^{-1}\bB^T\\
  & \textrm{(by Sherman-Morrison formula)}.
\end{align*}
Let $\bmu_b=\bB^T\bmu$ and denote $\bmu_b^*=\bB^T\hat{\bSig}^{-1}\bmu$,
then the second term simplifies to
\begin{align*}
  \bmu^T(\bI_L-2s\hat{\bSig})^{-1}\hat{\bSig}^{-1}\bmu&=(1-2s\hat{\sigma}^2)^{-1}\bmu^T\hat{\bSig}^{-1}\bmu-(1-2s\hat{\sigma}^2)^{-1}\bmu_b\left(\bD^{*2}-\frac{1-2s\hat{\sigma}^2}{2s}\bLam^{-1}\right)^{-1}\bmu_b^*\\
  &=(1-2s\hat{\sigma}^2)^{-1}\bmu^T\hat{\bSig}^{-1}\bmu+\sum_{i=1}^p\frac{2s\hat{\sigma}_i^2\mu_{bi}\mu^*_{bi}}{[1-2s(\hat{\sigma}_i^2d_i^2+\hat{\sigma}^2)](1-2s\hat{\sigma}^2)},
\end{align*}
where $\mu_{bi}$ and $\mu_{bi}^*$ are the $i$th entry of $\bmu_{b}$
and $\bmu^*_b$, respectively.

Finally, the Sherman-Morrison formula gives
$$\hat{\bSig}^{-1}= (\bB\bm{\Lambda}\bB^T+\hat{\sigma}^2\bI_L)^{-1} = \hat{\sigma}^{-2}\bI_L-\hat{\sigma}^{-2}\bB(\hat{\sigma}^2\bLam^{-1}+\bD^{*2})^{-1}\bB^T,$$
which can be used to simplify the third term as 
\begin{align*}
  \bmu^T\hat{\bSig}^{-1}\bmu=\frac{\bmu^T\bmu}{\hat{\sigma}^2}-\sum_{i=1}^p\frac{\hat{\sigma}^2_i\mu_{bi}^2}{\hat{\sigma}^2(\hat{\sigma}^2+\hat{\sigma}^2_id_i^2)}.
\end{align*}

The final expression of $\kappa_{\delta}(s)$ is obtained by taking the log-transform in the first term, and applying $\mu^*_{bi}=\mu_{bi}/(\hat{\sigma}^2+\hat{\sigma}_i^2d_i^2)$ in combining the two sums in the second and third term.
\end{proof}

\section{EI Criterion - Special Case}
\label{sec:EI criterion-trivial}

%\setcounter{result}{3}
%%
%\begin{result}\label{result:ei-trivial}
%  Suppose there is only one non-zero singular value $d_1$ in the SVD of the
%  response matrix $\bY$ of the dynamic computer simulator, then the
%  EI criterion (\ref{eq:ei-crit}) simplifies to
%\begin{align*}
%  \begin{aligned}
%  E[I(\bx)]=&\left[\delta_{\min}-\bxi^T\bxi+2\bxi^T\bb_1\hat{c}_1(\bx)-d_1^2\hat{c}^2_1(\bx)-%d_1^2\hat{\sigma}^2_1(\bx)\right]\left[\Phi(l_2(\bx))-\Phi(l_1(\bx))\right]\\
%  &+2\left[d_1^2\hat{c}_1(\bx)\hat{\sigma}_1(\bx)-\bxi^T\bb_1\hat{\sigma}_1(\bx)\right]
%\left[\phi(l_2(\bx))-\phi(l_1(\bx))\right]\\
%  &+d_1^2\hat{\sigma}_1^2(\bx)\left[l_2(\bx)\phi(l_2(\bx))-l_1(\bx)\phi(l_1(\bx))\right],
%  \end{aligned}
%\end{align*}
%when $(\bxi^T\bb_1)^2+d_1^2(\delta_{\min}-\bxi^T\bxi)>0$ and $E[I(\bx)]=0$
%otherwise, where $l_1(\bx)=(w_1-\hat{c}_1(\bx))/\hat{\sigma}_1(\bx)$, $l_2(\bx)=(w_2-\hat{c}%_1(\bx))/\hat{\sigma}_1(\bx)$, and
%{\color{black}
%\begin{align*}
%  \begin{aligned}
%  &w_1=\frac{\bxi^T\bb_1-\sqrt{(\bxi^T\bb_1)^2+d_1^2(\delta_{\min}-\bxi^T\bxi)}}{d_1^2}, \
%  w_2=\frac{\bxi^T\bb_1+\sqrt{(\bxi^T\bb_1)^2+d_1^2(\delta_{\min}-\bxi^T\bxi)}}{d_1^2}.
%  \end{aligned}
%\end{align*}
%}
%\end{result}

%------------------------

\begin{proof}[Proof of Result~4]
Let $\bY=d_1\bu_1\bv_1^T$, where $\bu_1\in\mathbb{R}^L$ and $\bv_1\in\mathbb{R}^N$ denote the left and right singular vectors of $\bY$. Therefore, the basis matrix (a vector in this case) is $\bB=\bb_1=d_1\bu_1$ and the coefficient matrix (vector) is $\bV^*=\bv_1$. As a result, $\by(\bx) = c_1(\bx)\bb_1$, and $\delta(\bx)  = \bxi^T\bxi - 2\bxi^T\bb_1c_1(\bx) + d_1^2c_1^2(\bx)$ (due to the orthogonality of $\bu_1$).
The integrand in the $E[I(\bx)]$ expression,  
$$\delta_{min}-\delta(\bx) = \delta_{\min}-\bxi^T\bxi+2\bxi^T\bb_1c_1(\bx)-d_1^2c_1^2(\bx),$$
is a quadratic function in $c_1(\bx)$ with a negative leading coefficient. Thus, $\delta_{min}-\delta(\bx) > 0$ in between the two roots (say, $w_1$ and $w_2$) if and only if the discriminant $(\bxi^T\bb_1)^2+d_1^2(\delta_{\min}-\bxi^T\bxi)> 0$. The roots of the corresponding quadratic equation will be
\begin{align*}
  \begin{aligned}
  &w_1=\frac{\bxi^T\bb_1-\sqrt{(\bxi^T\bb_1)^2+d_1^2(\delta_{\min}-\bxi^T\bxi)}}{d_1^2} \ \textrm{and} \
  w_2=\frac{\bxi^T\bb_1+\sqrt{(\bxi^T\bb_1)^2+d_1^2(\delta_{\min}-\bxi^T\bxi)}}{d_1^2}.
  \end{aligned}
\end{align*}
Subsequently, the EI criterion is
\begin{align*}
  E[I(\bx)]=&\int_{w_1}^{w_2}\left[\delta_{\min}
          -\bxi^T\bxi+2\bxi^T\bb_1c_1(\bx)-d_1^2c^2_1(\bx)
          \right]\pi(c_1(\bx)|\bv_1,\hbt_1)d c_1(\bx)\\
  =&\int_{w_1}^{w_2}\left[\delta_{\min}
          -\bxi^T\bxi+2\bxi^T\bb_1c_1(\bx)-d_1^2c_1^2(\bx)
          \right]
    \frac{1}{\sqrt{2\pi\hat{\sigma}^2_1(\bx)}}\exp\left[-\frac{(c_1(\bx)-\hat{c}_1(\bx))^2}{2\hat{\sigma}_1^2(\bx)}\right]d
    c_1(\bx)\\
    &(\textrm{substituting } c_1(\bx)= \hat{c}_1(\bx)+z\hat{\sigma}_1(\bx), \textrm{as a change of variable})\\
  =&\int_{l_1(\bx)}^{l_2(\bx)}\Big\{\delta_{\min}-\bxi^T\bxi+2\bxi^T\bb_1\big[\hat{\sigma}_1(\bx)z+\hat{c}_1(\bx)\big]-d_1^2\big[\hat{\sigma}_1(\bx)z+\hat{c}_1(\bx)\big]^2\Big\}
    \frac{1}{\sqrt{2\pi}}\exp\left[-\frac{z^2}{2}\right]d z\\
=&\left[\delta_{\min}-\bxi^T\bxi+2\bxi^T\bb_1\hat{c}_1(\bx)-d_1^2\hat{c}_1^2(\bx)\right]\int_{l_1(\bx)}^{l_2(\bx)}\phi(z)dz-d_1^2\hat{\sigma}^2_1(\bx)\int_{l_1(\bx)}^{l_2(\bx)}z^2\phi(z)d z\\
&+2\left[\bxi^T\bb_1\hat{\sigma}_1(\bx)-d_1^2\hat{c}_1(\bx)\hat{\sigma}_1(\bx)\right]\int_{l_1(\bx)}^{l_2(\bx)}z\phi(z)dz\\
=&\left[\delta_{\min}-\bxi^T\bxi+2\bxi^T\bb_1\hat{c}_1(\bx)-d_1^2\hat{c}^2_1(\bx)-d_1^2\hat{\sigma}^2_1(\bx)\right]\left[\Phi(l_2(\bx))-\Phi(l_1(\bx))\right]\\
  &+2\left[d_1^2\hat{c}_1(\bx)\hat{\sigma}_1(\bx)-\bxi^T\bb_1\hat{\sigma}_1(\bx)\right]\left[\phi(l_2(\bx))-\phi(l_1(\bx))\right]\\
  &+d_1^2\hat{\sigma}_1^2(\bx)\left[l_2(\bx)\phi(l_2(\bx))-l_1(\bx)\phi(l_1(\bx))\right],
\end{align*}
where the integrals in the last step have been simplified as per the properties of the standard normal distribution.
\end{proof}

\section{Convergence of the ESL2D Criterion}
\label{sec:appen-B}

%\setcounter{theorem}{0}

%\begin{theorem}
%Let $\by(\bx_i)$ be the time-series valued simulator response for $\bx_i \in \Omega=[0,1]^q$, for $i=1,...,N$, and $\bxi$ be the pre-specified target output. If $\by(\bx)$ is emulated via the SVD-based GP model (as in Section~\ref{sec:svd-gpm}), and the discrepancy $\delta(\bx)=\|\bxi-\by(\bx)\|^2_2$ is a scalar valued continuous function in $\bx$, then the ESL2D optimal solution $\hat{\bx}^*$ in (\ref{eq:inv-ESL2D}) converges in probability to the true solution $\bx^*$ in (\ref{eq:ls-sol}), as $N \rightarrow \infty$, i.e.,
%%
%\begin{align*} 
%  \underset{\bx\in\Omega}{\mathrm{argmin}}\: \tE\big[\delta(\bx)\big|\bY_N\big]\quad  
%\overset{p}{\longrightarrow}  \quad \underset{\bx\in\Omega}{\mathrm{argmin}}\: \delta(\bx). 
%\end{align*}
%%
%\end{theorem}
%------------------------

\begin{proof}[Proof of Theorem~1] The objective is to show
\begin{align*} 
  \underset{\bx\in\Omega}{\mathrm{argmin}}\: \tE\big[\delta(\bx)\big|\bY\big]\quad  
\overset{p}{\longrightarrow}  \quad \underset{\bx\in\Omega}{\mathrm{argmin}}\: \delta(\bx), 
\end{align*}
as $N \rightarrow \infty$. Note that most of the variables and parameter estimates depend on $N$, however, for keeping the notations simpler, we have not shown explicit dependency on $N$. For eample, we use $\bY$ and $\bU$ instead of $\bY_N$ and $\bU_N$ respectively. 

Since $\delta(\bx)$ is a continuous function of $\bx \in \Omega= [0,1]^q$, it is sufficient to show that 
\begin{align*}
 \underset{\bx\in\Omega}{\sup}\left|\tE[\delta(\bx)|\bY]-\delta(\bx)\right|\xrightarrow{p} 0.
\end{align*}
Note that one can also prove the continuity of $\delta(\bx)$ and sufficiency of the supremum condition (above), however, keeping the space constraint in mind, we have skipped this part of the proof. From the proof of Result~1 in Appendix~A, 
$$ E[\delta(\bx)|\bY] = \|\bxi-\hat{\by}(\bx)\|^2+ tr\left(\bLam(\bV)\bB^T\bB\right)+\hat{\sigma}^2L,$$
and 
\begin{align*}
  \|\bxi-\hat{\by}(\bx)\|_2^2&=\|\bxi-\by(\bx)\|_2^2+\|\by(\bx)-\hat{\by}(\bx)\|^2_2+2\langle\bxi-\by(\bx),\by(\bx)-\hat{\by}(\bx)\rangle.
\end{align*}
We further assume that there is no residual term in the SVD of $\bY$, that is, $\hat{\sigma}^2=0$. Therefore,
\begin{align*}
 \tE[\delta(\bx)|\bY]-\delta(\bx) = \|\by(\bx)-\hat{\by}(\bx)\|^2_2+2\langle\bxi-\by(\bx),\by(\bx)-\hat{\by}(\bx)\rangle + tr\left(\bLam(\bV)\bD^2\right).
\end{align*}
Subsequently, it is enough to show that 
\begin{itemize}
 \item[(i)] $\underset{\bx\in\Omega}{\sup}\left|\|\by(\bx)-\hat{\by}(\bx)\|_2^2+2\langle\bxi-\by(\bx),\by(\bx)-\hat{\by}(\bx)\rangle_2\right|\xrightarrow{p}  0,$ 
 \item[(ii)] $\underset{\bx\in\Omega}{\sup}\:\tr(\bLam(\bV)\bD^2)\xrightarrow{p} 0.$ 
\end{itemize}
For part (i), we have to show that 
$$\underset{\bx\in\Omega}{\sup}\|\by(\bx)-\hat{\by}(\bx)\|_2^2+ \underset{\bx\in\Omega}{\sup} \|\bxi-\by(\bx)\|_2 \|\by(\bx)-\hat{\by}(\bx)\|_2\xrightarrow{p}  0.$$
Assuming that the supremum of the true (unobserved) discrepancy over the input space is finite, i.e., $\underset{\bx\in\Omega}{\sup} \|\bxi-\by(\bx)\|_2 < \infty$, a sufficient condition for proving part (i) is to show that
$$\underset{\bx\in\Omega}{\sup}\|\by(\bx)-\hat{\by}(\bx)\|_2^2\xrightarrow{p}  0.$$
Since the left singular vector matrix $\bU$ is an $L\times L$ orthonormal matrix,  
\begin{align*} 
  \begin{aligned}
  \|\by(\bx)-\hat{\by}(\bx)\|^2_2
  &=\|\bU^T\by(\bx)-\bD\hat{\bc}(\bx|\bV)\|_2^2  
  =\sum_{i=1}^L\left(\bu_{i}^T\by(\bx)-d_i\hat{c}(\bx|\bv_{i})\right)^2\\ 
  &=\sum_{i=1}^{L'}d_{i}^2\left(w_{i}(\bx)-\hat{c}_i(\bx|\bv_{i})\right)^2+\sum^L_{i=L'+1}\left(\bu_{i}^T\by(\bx)\right)^2,
  \end{aligned}  
\end{align*}
where $L'=\max\{i:1\leq i\leq L,d_{i}>0\}$ and $w_{i}(\bx)=d_{i}^{-1}\bu^T_{i}\by(\bx)$. Using some properties of reproducing kernel Hilbert space (RKHS), denoted by $\mathcal{N}_{K}(\Omega)$, and the existence of $2m$ order of continuous derivatives of the correlation kernel $K$ (see Theorem~11.13 of \cite{wendland2004} for details), one can show that 
\begin{align*}
  \underset{\bx\in\Omega}{\sup}|w_{i}(\bx)-\hat{c}_i(\bx|\bv_{i})|\leq C_{K}h_{\bX,\Omega}^m\|w_{i}\|_{\mathcal{N}_{K}(\Omega)},
\end{align*}
where $C_K>0$ is a constant which depends only upon the correlation kernel $K$, $h_{\bX,\Omega}$ is the fill distance of design $\bX=\{\bx_1,\dots,\bx_N\}$ on a bounded
  domain $\Omega$, defined by
  \begin{align*}
    h_{\bX,\Omega}=\sup_{\bx\in\Omega}\min_{1\leq i\leq N}\|\bx-\bx_i\|_2,
  \end{align*}
and $\|w_{i}\|_{\mathcal{N}_{K}(\Omega)}$ is the RKHS norm of $w_i(\bx)$, given by,
\begin{align*}
  \|w_{i}\|_{\mathcal{N}_{K}(\Omega)}&=d_{i}^{-1}\|\bu_i^T\by(\bx)\|_{\mathcal{N}_{K}(\Omega)}\leq d_{i}^{-1}\|\by(\bx)\|_{\mathcal{N}_{K}(\Omega)}.
\end{align*}
Note that the Gaussian correlation kernel is infinitely differentiable and satisfies all the RKHS properties. As a result, we get
\begin{align*} 
  \begin{aligned}
 \underset{\bx\in\Omega}{\sup}\, \sum_{i=1}^{L'}d_{i}^2\left(w_{i}(\bx)-\hat{c}_i(\bx|\bv_{i})\right)^2  &\le   \sum_{i=1}^{L'} d_{i}^2\, \underset{\bx\in\Omega}{\sup}\, \left(w_{i}(\bx)-\hat{c}_i(\bx|\bv_{i})\right)^2\\
 &\le L' C_{K}^2 h_{\bX,\Omega}^{2m} \|\by(\bx)\|_{\mathcal{N}_{K}(\Omega)}.
  \end{aligned}  
\end{align*}
Since $\|\by(\bx)\|_{\mathcal{N}_{K}(\Omega)} = O(1)$, the convergence of fill-distance criterion \citep{wendland2004} implies 
$$\underset{\bx\in\Omega}{\sup}\, \sum_{i=1}^{L'}d_{i}^2\left(w_{i}(\bx)-\hat{c}_i(\bx|\bv_{i})\right)^2 \overset{p}{\longrightarrow} 0.$$

For the second part of $\underset{\bx\in\Omega}{\sup}\|\by(\bx)-\hat{\by}(\bx)\|_2^2$, the uniform continuity of $\by(\bx)$ on $\Omega$ ensures that for any $\epsilon>0$ there exists an $\eta>0$ such that
\begin{align*}
P\left(\sup_{\bx\in\Omega}\left(\bu_{i}^T\by(\bx)\right)^2>\epsilon\right)\leq
P(h_{\bX,\Omega}>\delta)\rightarrow 0,
\end{align*}
as $N\rightarrow\infty$ (note that $\bu_i$ and $\bX$ depends on $N$), which implies that
$$\underset{\bx\in\Omega}{\sup} \sum^L_{i=L'+1}\big(\bu_{i}^T\by(\bx)\big)^2=o_p(1).$$

For part (ii), recall that  $\bLam(\bV)$ is an $L\times L$ diagonal matrix with the $i$-th entry being $\hat{\sigma}^2_i$ (as in (\ref{eq:coef-ps2})). Thus,
\begin{align*}
   \underset{\bx\in\Omega}{\sup}\:\tr(\bLam(\bV)\bD^2)= \underset{\bx\in\Omega}{\sup}\:\sum_{i=1}^L d_i^2 \hat{\sigma}^2_i .
\end{align*}
From Theorem~11.13 of \cite{wendland2004}, the properties of RKHS and the existence of $2m$ order of continuous derivatives of the correlation kernel implies that 
\begin{align*}
  \sup_{\bx\in\Omega}\left(1-\bk^T(\bx)\bK^{-1}\bk(\bx)\right)\leq
  C_K^2 h^{2m}_{\bX,\Omega},
\end{align*}
and 
\begin{align*}
\bv_{i}^T\bK^{-1}\bv_{i}\leq
\|w_{i}\|_{\mathcal{N}_{K}(\Omega)}^2\leq d_{i}^{-2} \|\by(\bx)\|^2_{\mathcal{N}_{K}(\Omega)}.
\end{align*}
Again, the boundedness of the norm, $\|\by(\bx)\|^2_{\mathcal{N}_{K}(\Omega)} = O(1)$, and the convergence of fill distance criterion, $h_{\bX,\Omega}$, implies $\underset{\bx\in\Omega}{\sup}\ \tr(\bLam(\bV)\bD^2) \xrightarrow{p}0$ as $N \rightarrow \infty$.
\end{proof}

\bibliographystyle{chicago}
\bibliography{l2inv}

\clearpage

\fontsize{12}{14pt plus.8pt minus .6pt}\selectfont \vspace{0.1pc}
\centerline{\LARGE\bf Supplementary Material}\vspace{0.2in}

\renewcommand{\thesection}{S.\arabic{section}}
\renewcommand{\theequation}{S.\arabic{equation}}
\renewcommand{\thefigure}{S.\arabic{figure}}
\renewcommand{\thetable}{S.\arabic{table}}

\setcounter{section}{0}
\setcounter{figure}{0}

% \section{Empirical Study on Saddlepoint Approximation for the EI Criterion}
% \label{sec:s-saddleapprox}
In this supplementary material, we provide an empirical study for the
performance comparison of the saddlepoint approximation-based expected improvement (saEI) criterion and a Monte Carlo (MC) method for approximating the EI criterion. We also compare two methods (referred to as the naive method and the proposed ESL2D method) for extracting the solution of the inverse problem. It is demonstrated that the saEI criterion gives  high accuracy in much less computational time as compared to the MC integration for evaluating (\ref{eq:ei-crit}). The  proposed method via ESL2D also significantly outperforms the naive approach in extracting the best estimate of the inverse solution.

%We also provide a brief summary of the performance of the saddlepoint
%approximation compared with the Monte Carlo (MC) approximation in the
%examples studied in Section \ref{art-sec:sim-study} of the article.
\section{The accuracy of the saEI criterion}

We present two detailed examples to demonstrate the accuracy of the proposed saEI criterion.  In Example S1, the exact formula for the EI criterion  (\ref{eq:ei-crit}) is available while in Example S2 in which such a formula is unavailable, the Monte Carlo (MC) approximations are used as a bench mark.  For simplicity, we considered simulators with only one-dimensional inputs.

\begin{figure}[h!]
  \centering
  \includegraphics[height=4in]{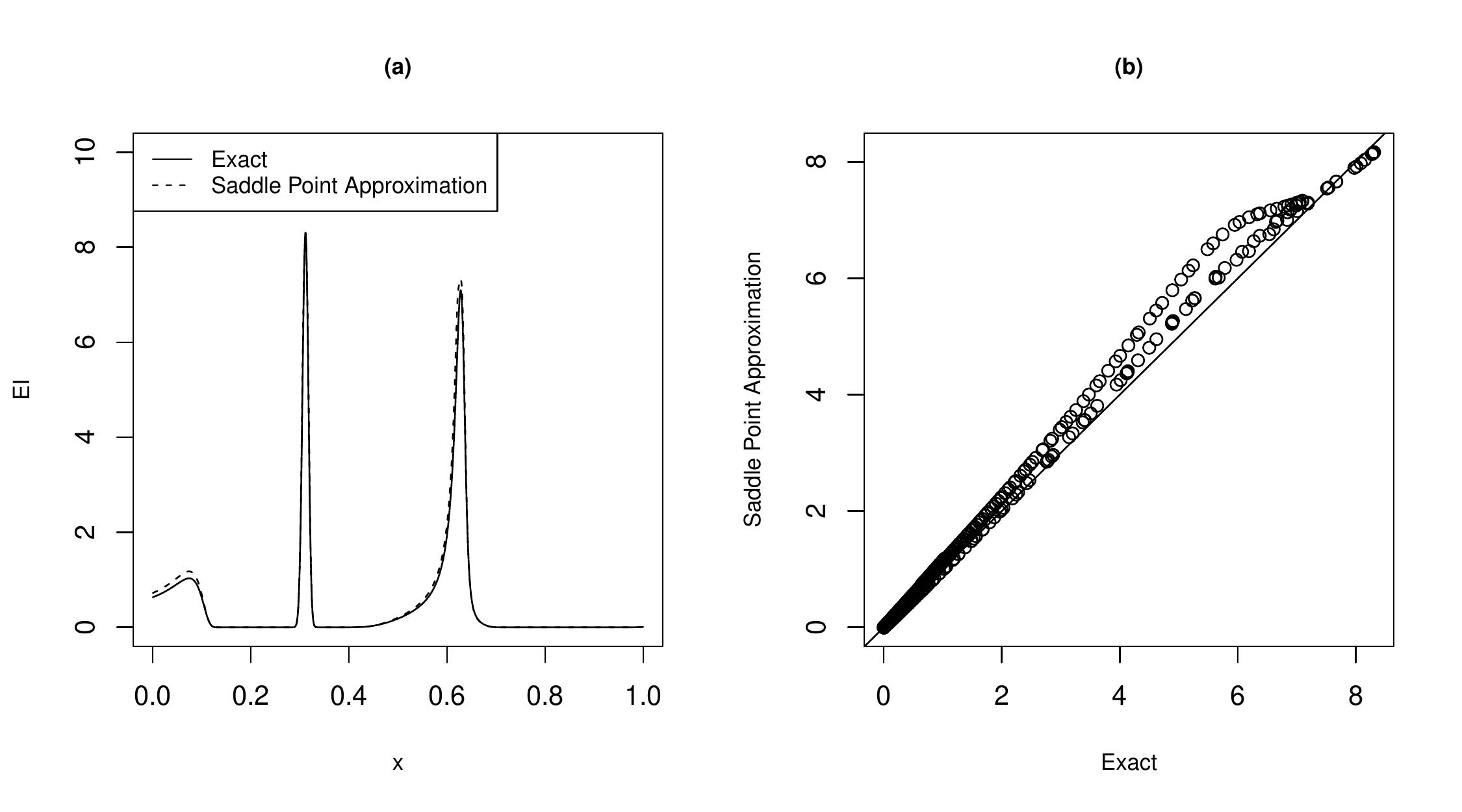}
  \caption{The exact EI (as in Result~4) and saEI
    criteria (as in (\ref{eq:saddle-approx-pos}) - 
(\ref{eq:saddle-approx-neg})) for Example S1. Panel (a): The curves of the exact and
    saEI criterion values on the design domain
    $[0,1]$. Panel (b): The scatter plot of the exact values against saEI criterion values.}
\label{fig:exactvapprox}
\end{figure}

\textbf{Example S1}: Consider the synthesized computer simulator
$y_t(x)=g(t)g(2x+0.5)$, where $g(t)$ is the function studied by
\cite{gramacy2012cases}, i.e.,
\begin{align*}
  g(t)=\frac{\sin(10\pi t)}{2t}+(t-1)^4.
\end{align*}
For the simulator $y_t(x)$, the design input $x\in[0,1]$ and the time
$t\in[0.5,2.5]$ is on a 200-point equidistant time grid. The mechanism
of generating the field observation $\bxi$ is exactly the same as that
in Example~1 in Section \ref{sec:illu-ex-cont}. First
we randomly generate an $x^*\in[0,1]$. The field
observation $\bxi=[\xi_{t_1},\dots,\xi_{t_{200}}]^T$ is generated by
$\xi_{t_i}=y_{t_i}(x^*)+e_i$, where $e_i$'s are Gaussian white noises
with mean 0 and variance 0.03522 to produce a signal to noise ratio
50. The initial design is a 5-point random Latin hypercube design (LHD).
The corresponding data is fit via the SVD-based GP model. In this example, the first singular value explains
100\% of the variation in the response matrix of the design
set. Therefore, $p=1$ and $\hat{\sigma}^2=0$, which implies that
the EI criterion (\ref{eq:ei-crit})  has
explicit formula as given in Result 4. We evaluate the exact value of the EI criterion and the saEI criterion in (\ref{eq:saddle-approx-pos}) - 
(\ref{eq:saddle-approx-neg}) on  a random LHD of  $2,000$
points in $[0,1]$. The
comparison of the exact and approximate values is provided in Figure
\ref{fig:exactvapprox} which indicates that they  are nearly identical. 
%,natheight=400

%=========================================

\begin{figure}[h!]
  \centering
  \includegraphics[height=4in]{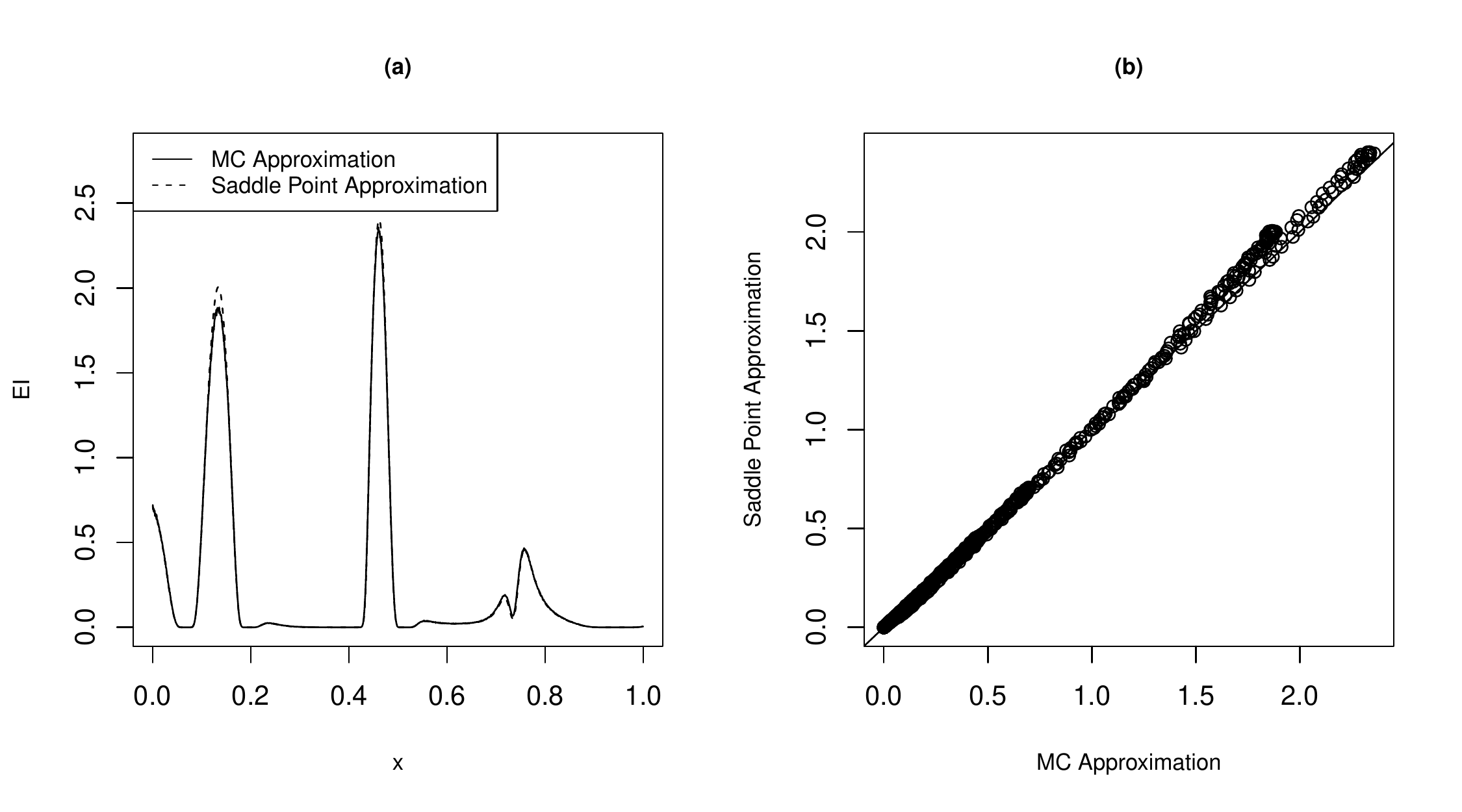}
  \caption{The MC approximation of the EI criterion and the saEI
    criterion in Example S2. (a) the curves of the MC approximation and the  saEI criterion values on the design domain
    $[0,1]$, (b) the scatter plot of the MC against the saEI criterion values.}
\label{fig:mcvapprox}
\end{figure}

\textbf{Example S2}: We revisit  Example~1 in Section \ref{sec:illu-ex-cont} of the main paper. In this example, the number of significant singular values is $p=4$. Therefore, no explicit formula for the EI
criterion is available. As a bench mark, we use Monte Carlo (MC)
approximation with $50,000$ random samples, which shall yield sufficiently
accurate approximation of  (\ref{eq:ei-crit}). The saEI criterion values for selecting the first follow-up design point
are computed on a random LHD of $2,000$ points in $[0,1]$.  Figure \ref{fig:mcvapprox} provides the approximation of 
  (\ref{eq:ei-crit}) via MC approximation and saEI criterion values. 
  The figure reveals that the two approximation methods provide very close values. The MC approximation takes 1621.24 seconds to evaluate
the 2,000 EI values  on our workstation (AMD
A10-6700 3.70 GHz, 8 GB RAM, 64-bit Windows 10). In contrast, the
saEI method takes 0.06 seconds.

\begin{figure}[h!]
  \centering
  \includegraphics[height=3in]{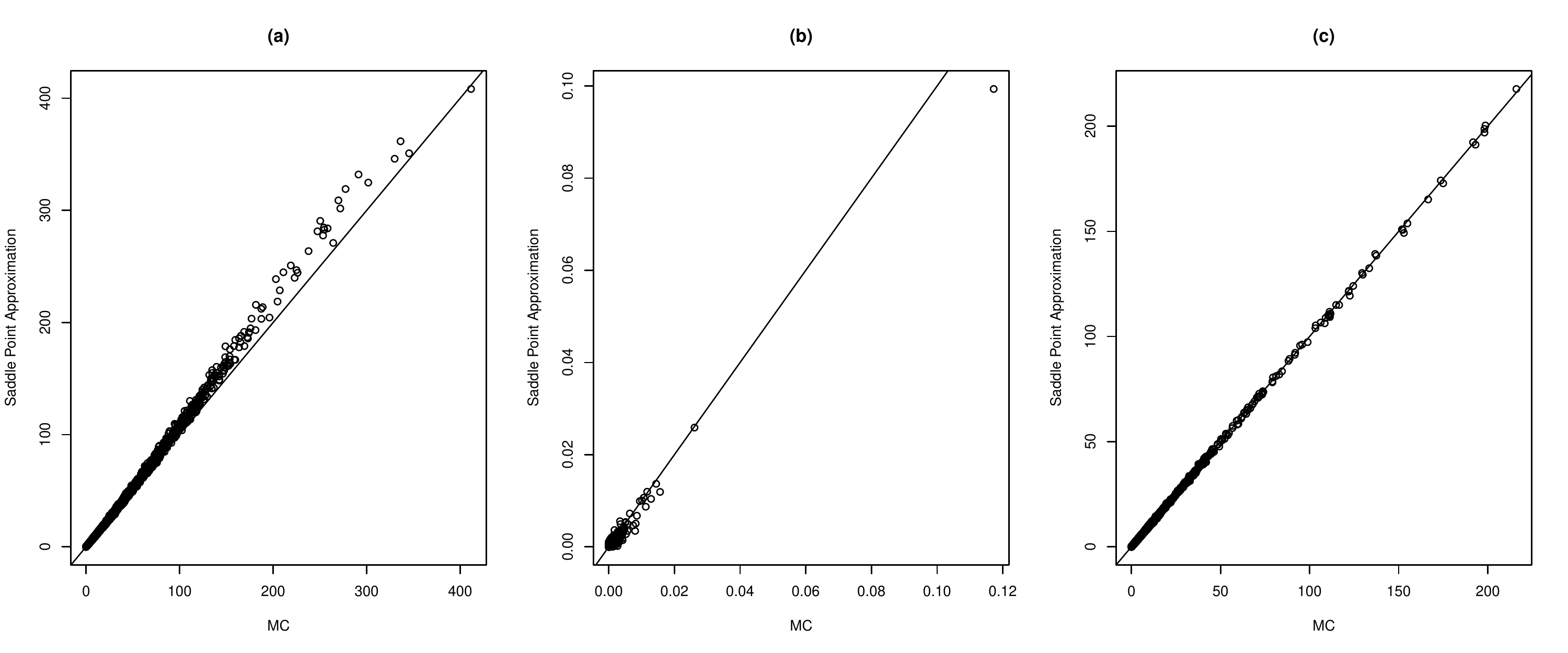}
  \caption{The MC approximation of the EI criterion and the saEI
    criterion for selecting the first follow-up design point in (a)
    Example 2, (b) Example 3, (c) the real application in Section
    \ref{sec:sim-study}.}
\label{fig:mcvappexam}
\end{figure}

We have also conducted the comparison between the MC approximation of the EI criterion and the
saEI criterion in Examples 2, 3 and
the real application studied in Section \ref{sec:sim-study}. For reasons of saving space, here we only provide the approximate values of the EI
criterion for selecting the first follow-up point in each example. To be consistent with
the simulation setup in Section \ref{sec:sim-study}, the
saEI criterion values are evaluated on sets with
$2000q$ points, where $q$ is the number of input variables in the
computer simulators. The MC approximation uses $50,000$
random samples for each input in the sets. The scatter plots
of the MC approximation values and the saEI criterion values
for the three examples are provided in Figure \ref{fig:mcvappexam}. The figure again shows that the MC approximation and the saEI criterion provide similar approximate values for the
EI criterion on the set in all the three examples. In terms
of computational time, in all the three examples, the MC approximation
takes more than 1,000 seconds, in contrast, the saEI criterion only takes less than 1 second.

% The execution time of the MC and the saddlepoint
% approximation is provided in Table \ref{tab:mcvappexam}.
% \begin{table}[h!]
%   \centering
%   \begin{tabular}{|l|lll|}
% \hline
%     &Example 2&Example 3&Real Application\\
% \hline
% MC Approximation&5744.23&9832.93&1307.07\\
% Saddlepoint Approximation&0.34&0.46&0.37\\
% \hline
%   \end{tabular}
%   \caption{Execution time (in seconds) of the MC and the saddlepoint
%     approximation in Examples 2 and 3 and the real application in Section
%     \ref{art-sec:sim-study}.}
%   \label{tab:mcvappexam}
% \end{table} 

\section{Comparison of the naive and ESL2D method}

\begin{figure}[h!]\centering
 \includegraphics[height=3.35in]{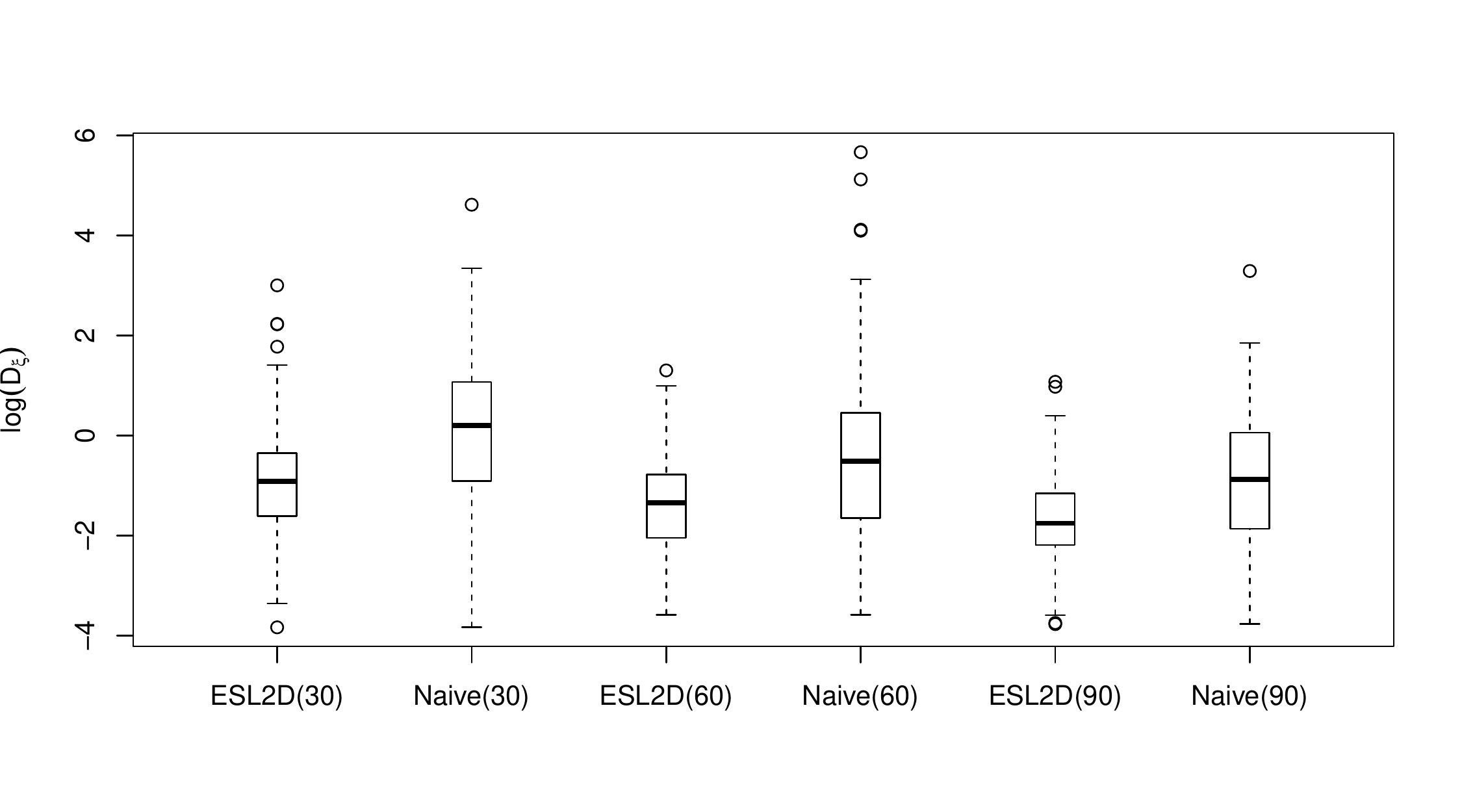}\vspace{-0.8in}
 \includegraphics[height=3.35in]{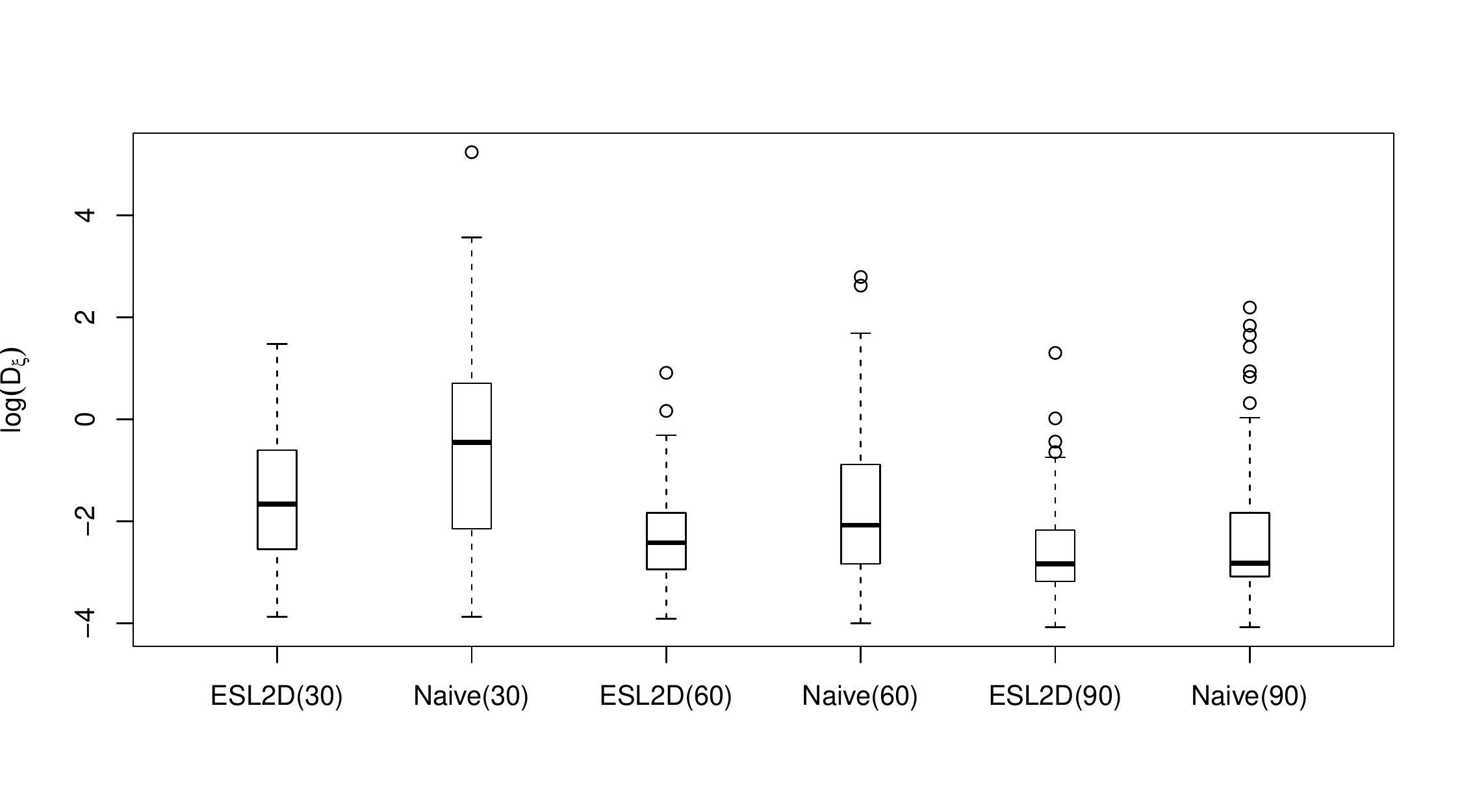}
 \caption{Example~2 \citep{harari2014convex}. Distribution of $\log(D_\xi)$ values from the final fit. Top panel: Space-filling design, bottom panel: sequential design with $n_0=2n/3$. The number in the parenthesis is the run size $n$.}
 \label{fig:supp:harari}
\end{figure}

\begin{figure}[h!]\centering
 \includegraphics[height=3.35in]{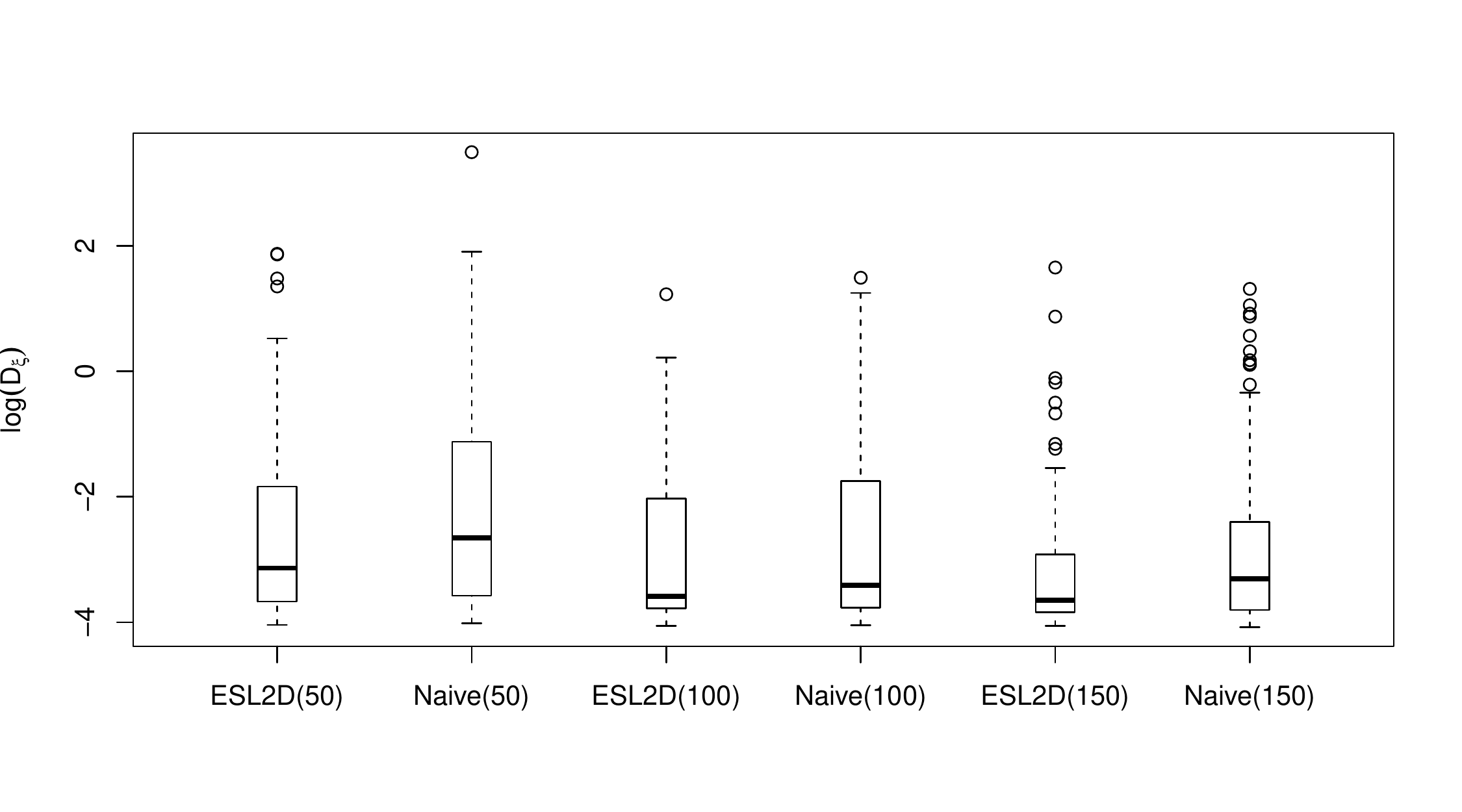}\vspace{-0.8in}
 \includegraphics[height=3.35in]{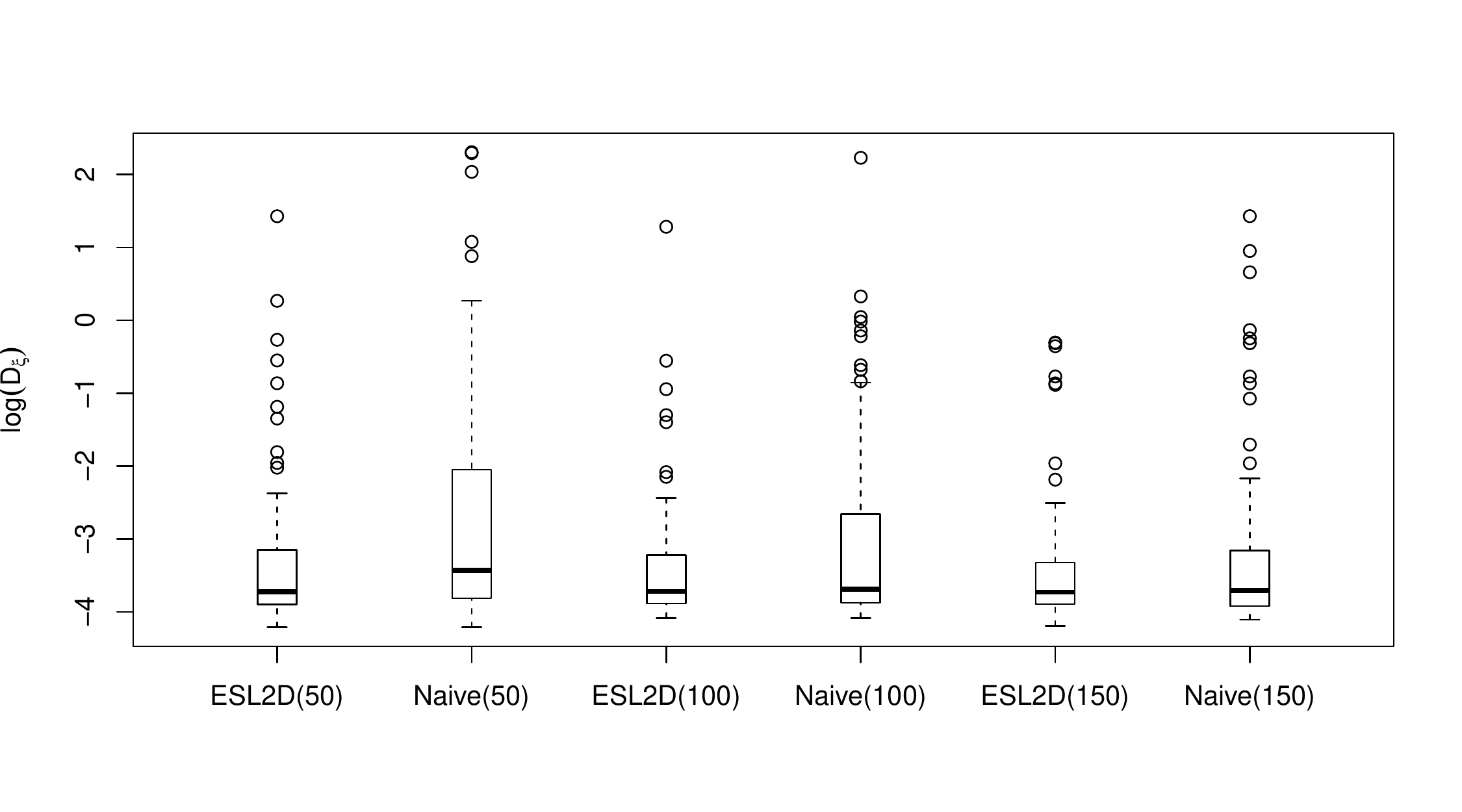}
 \caption{Example~3 \citep{bliznyuk2008bayesian}.  Distribution of $\log(D_\xi)$ values from the final fit. Top panel: Space-filling design, bottom panel: sequential design with $n_0=2n/3$. The number in the parenthesis is the run size $n$.}
  \label{fig:supp:bliz}
\end{figure}

The objective here is to compare the accuracy of the extracted inverse solution. Thus, we compare $\log(D_\xi)$ values of the final inverse solutions obtained via the naive method ($\tilde{\bx}^*$) and the ESL2D method ($\hat{\bx}^*$). We used test simulators from Examples~2 and 3 from the main paper for this comparison. Each boxplot shown in Figures~\ref{fig:supp:harari} - \ref{fig:supp:bliz} represents the distribution of $\log(D_\xi)$ over 100 simulations. We extract the inverse solution in two scenarios: (a) used a  space-filling Latin hypercube design in the input space to fit the emulator via an SVD-based GP model and then extracted the solution, and (b) used the proposed sequential design approach with the saEI criterion and then extract the solution from the final fit. The first method (space-filling design study) simply provides values for the reference comparison. 

For both scenarios, we used $n = 10q, 20q$ and $30q$, where $q$ is the input dimension. For the sequential framework, we used initial designs of size $n_0=2n/3$, and the remaining $n-n_0$ follow-up points were obtained one at-a-time by maximizing the saEI criterion. For each simulation, we randomly generated the target response, the training design points, the candidate set for selecting follow-up points, and the candidate set for extracting the inverse solutions. Note that the signal part of the target response is held fixed over all the simulations in the main part of the article, however, in this section, we let it vary to account for more uncertainty.

Figure~\ref{fig:supp:harari} presents the simulation results for Example~2 of the main paper, which is a three-dimensional test simulator from \cite{harari2014convex}. It is clear from the figure that the proposed ESL2D approach gives more accurate estimate of the inverse solutions as compared to the naive approach. As expected, the accuracy increases with $n$, and the relative superiority of ELS2D over the naive approach decreases with $n$.

Figure~\ref{fig:supp:bliz} shows the simulation results for the test simulator in Example~3 of the main paper. This is a five-dimensional test simulator from \cite{bliznyuk2008bayesian}. The observations are very similar as in the previous example.

\end{document}